\documentclass[useamsfonts]{pasj01}
\draft

\begin{document}
\Received{yyyy/mm/dd}
\Accepted{yyyy/mm/dd}
\Published{yyyy/mm/dd}

\title{A Formulation of Consistent Particle Hydrodynamics in Strong Form}

\author{Satoko \textsc{Yamamoto}\altaffilmark{1,2,3}}%
\altaffiltext{1}{Department of Earth \& Planetary Sciences, Tokyo Institute of Technology, 2-12-1, Ookayama, Meguro-ku, Tokyo 152-8550, Japan}
\altaffiltext{2}{RIKEN Advanced Institute for Computational Science, 7-1-26, Minatojima-minamimachi, Chuo-ku, Kobe, Hyogo 650-0047, Japan}
\altaffiltext{3}{Department of Planetology, Kobe University, 1-1, Rokkodaicho, Nada-ku, Kobe, Hyogo 650-0013, Japan}
\email{yamamoto.s.an@geo.titech.ac.jp}

\author{Junichiro \textsc{Makino},\altaffilmark{3,2,1}}



\maketitle

\begin{abstract}
In fluid dynamical simulations in astrophysics, large deformations are
common and surface tracking is sometimes necessary. Smoothed Particle
Hydrodynamics (SPH) method has been used in many of such simulations.
Recently, however, it has been shown that SPH cannot handle contact
discontinuities or free surfaces accurately. There are several reasons for
this problem. The first one is that SPH requires that the density is
continuous and differentiable. The second one is that SPH does not have
the consistency, and thus the accuracy is zeroth order in space. In
addition, we cannot express accurate boundary conditions with SPH. In this
paper, we propose a novel, high-order scheme for particle-based
hydrodynamics of compressible fluid. Our method is based on kernel-weighted
high-order fitting polynomial for intensive variables. With this approach,
we can construct a scheme which solves all of the
three problems described above. For shock capturing, we use a
tensor form of von-Neumann-Richtmyer artificial viscosity. We have applied
our method to many test problems and obtained excellent result.
Our method is not conservative, since particles do not have
mass or energy, but only their densities. However, because of
the Lagrangian nature of our scheme, the violation of the
conservation laws turned out to be small. We name this
method Consistent Particle Hydrodynamics in Strong Form (CPHSF).
\end{abstract}

\section{Introduction}
\label{sec:intro}
In fluid dynamical simulations in astrophysics, large deformations are
common and surface tracking is sometimes necessary. Mesh-free methods, in
which particles move following the motion of fluid, are very useful for
such simulations. In particular, Smoothed Particle Hydrodynamics (SPH;
\cite{1977AJ.....82.1013L}, \cite{1977MNRAS.181..375G}) has been
widely used in astrophysics and also in computer-aided engineering.

SPH is one of Lagrangian methods. In SPH, we assume that the fluid equation
can be expressed by interactions between fluid particles. Therefore, SPH is
not only suitable for simulation of large deformations, but also can
satisfy the conservation laws.

Recently, however, it has become known that SPH has several difficulties.
For example, it cannot handle contact discontinuities (e.g.
\cite{2003MNRAS.345..429O}, \cite{2007MNRAS.380..963A}) or fluid surfaces.
In our opinion, the standard formation of SPH has the following three
problems, 1) The density distribution must be differentiable (e.g.
\cite{2013ApJ...768...44S}). Hence, SPH cannot handle the contact
discontinuity properly. 2) Since the approximation of quantities in
SPH is of zeroth order in space, SPH does not have the consistency to the
original partial differential equation (e.g. \cite{liu1995reproducing}).
According to the Lax equivalence theorem (\cite{lax1956survey}), a method,
which does not have the consistency, does not converge to the original partial
differential equation in the limit of the infinite resolution. 3) There is
no mathematically sound way to specify boundary conditions in SPH, except
for the mirror boundary condition. Traditionally, fixed particles have been
used to express reflecting boundaries such as walls and bottom of a well.
They are necessary because SPH cannot express a sharp cutoff in the density
distribution. However, there is no way to let these fixed
particles change their physical quantities correctly. Thus, smoothed
estimate of physical quantities of particles near the boundary contains
large errors.

To solve problem one, \citet{2001MNRAS.323..743R},
\citet{2003physics...3112O}, \citet{2013ApJ...768...44S} and
\citet{2015PASJ...67...37Y} proposed modified formulation of SPH in which
the differentiability of the density is not required. Their methods can
handle the density discontinuity better than standard SPH (hereafter SSPH)
does.

In previous studies, high-order scheme have been proposed as the solution
to problem two. \citet{1999IJNME..44.1115D} formulated Moving Least Squares
Particles Hydrodynamics (MLSPH) based on the Moving Least Square (MLS)
technique. In Reproducing Kernel Particle Method (RKPM;
\cite{liu1995reproducing}), the formulation is not MLS, but is similar.
These methods were applied to inviscid fluid dynamics simulations. However,
they have not been applied to large deformations. Corrective SPH Method
(CSPM; \cite{chen1999improvement}, \cite{chen1999completeness},
\cite{chen1999corrective}, \cite{chen2000generalized}) is based on the
Taylor expansion and Method of Weighted Residuals (MWR). They calculated
Burgers equation, conduction of heat, linear elastodynamics and others with
CPHM. In addition, \citet{zhang2004modified} modified this scheme and they
called it Modified SPH (MSPH). MSPH was applied to the elastic wave and
the diffusion equation.
For non-compressible fluid, \citet{tamai2013development} introduced
high-order formulation into Moving Particle Semi-implicit method
(\cite{1996NSE...123..421K}).
In addition, Finite Particle Method (FPM; \cite{liu2005modeling}) was
developed using MWR to handle viscous fluid. They calculated dam
break test with FPM.
Corrected SPH (CSPH) in \citet{staroszczyk2010simulation} successfully
handled the dam break test of inviscid fluid by re-fitting the density
frequently. The frequent re-fitting generates numerical viscosity, and
particles move to reduce the number-density deviation. Thus, large
viscosity occurs in their simulation. In other word, if fluid particles do
not move following fluid line to prevent the number-density deviation, large
numerical viscosity is induced.

\begin{figure}[ht]
	\begin{center}
	\includegraphics[width=12cm,clip]{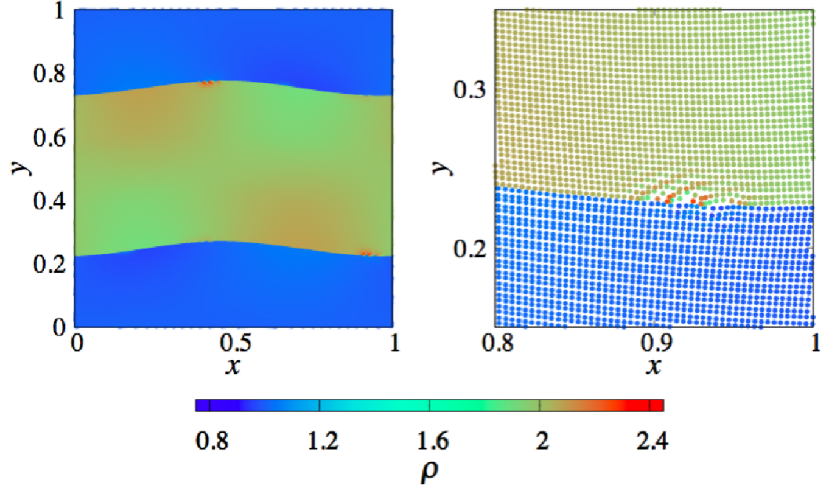} 
	\end{center}
	\caption{The result of a simulation of the Kelvin-Helmholtz
		instability with our high-order
		method at $t = 0.23\tau_{\rm KH}$. The right panel is the
		enlarge image of low-density region of the left.}
	\label{fig:khi_noise}
\end{figure}

There seems to be no high-order scheme without large numerical
viscosity for inviscid fluid tested with the calculation of large
deformations. \citet{2016arXiv160500725F} argued that it is difficult to
handle large deformations with a high-order scheme.
Figure {\ref{fig:khi_noise}} shows the result of a
simulation of the Kelvin-Helmholtz instability performed using the
high-order mesh-free method presented in section \ref{sec:form_norm}. The
simulation time is $t = 0.23\tau_{\rm KH}$, where $\tau_{\rm KH}$ is the
time scale of the Kelvin-Helmholtz instability. We can see large deviations
of the number-density of particles. These deviations are the result of the
fact that each particle move following the fluid motion at its location
accurately. Small-scale vortex can easily generate highly disordered
distribution of particles. This problem is mathematically same as the
generation of large density fluctuation in cold Keplerian disk (e.g.
\cite{2002ApJ...569..501I}).
From the above, it is necessary to rearrange particles, when the large
number-density deviation occurs.

One potential problem of high-order method is that they do not completely
satisfy the conservation law because particles do not have extensive variables
(\cite{fang2009improved}, \cite{2016arXiv160500725F}). To construct a
high-order mesh-free method which satisfies the conservation laws, we must
define fluid particles which have extensive variables. Hence, each particle
must also have its volume. This volume has to be represented by some
physical shape of the particle. If the mass of a particle
is constant, its shape of the particle has to change following the fluid
deformation. Consider an initially spherical particle in velocity field with a
uniform shear. It will become elongated very soon, and thus numerical scheme
would break down. Therefore, no high-order method can satisfy the
conservation law completely. However, we found that the error of the
conservation law is very small for high-order methods.

Concerning problem three, we can apply boundary conditions in mathematically
meaningful and well-defined way because we approximate the partial
differential equation using fitting polynomial for intensive variables in
high-order mesh-free method.

We call our method Consistent Particle Hydrodynamics in Strong Form
(CPHSF). We performed several numerical tests, and results were excellent.

In the rest of this paper, we present CPHSF (section \ref{sec:form}), and
report the results of numerical tests (section \ref{sec:test}). Finally, we
discuss and summarize our study (section \ref{sec:sumanddis}).

\section{The CPHSF scheme}
\label{sec:form}
As described in introduction, many different high-order approximation
methods have been proposed in previous studies. In the following, we call these
approximation methods "shape functions", since they can be written as
\begin{eqnarray}
	\hat{f}_{x,\dots,y,\dots,z,\dots}\large|_{\boldsymbol{r} = \boldsymbol{r}_i}
	\equiv \sum_j f_j \mathcal{W}_j(\boldsymbol{r}_j - \boldsymbol{r}),
\end{eqnarray}
where $\hat{f}_{x,\dots,y,\dots,z,\dots}$ is the approximation of a spacial partial
derivative of function $f$ with respect to variables
${x,\dots,y,\dots,z,\dots}$ at position $\boldsymbol{r}$, $f_j$ is the
value of function $f$ at the location of particle $j$, $\boldsymbol{r}_j$, and
$\mathcal{W}_j$ is the "shape function" of particle $j$. These shape function
can be classified by whether or not they are defined as the minimum of the
L2 norm of the residual of the fitting. We call shape functions based on
the minimization of L2 norm type 1 and others type 2.

\subsection{Type 1 shape functions}
\label{sec:form_norm}
In this section, we consider an approximation of function $f$,
$\hat{f}(\boldsymbol{r})$, which minimizes
 \begin{eqnarray}
	\label{eq:org,error,normal}
	\epsilon
	= \sum_j \widetilde{W}_{ij}
	\left[\hat{f}(\boldsymbol{r}_j) - f_j\right]^2.
 \end{eqnarray}
We assume $f(\boldsymbol{r})$ is of $C^n$ class, i.e. $f(\boldsymbol{r})$
is $n$-times differentiable. Here, $\widetilde{W}_{ij}$ is the weight of particle $j$
for approximation at particle $i$.

To derive shape functions in the way similar to those in \citet{staroszczyk2010simulation}
and \citet{tamai2013development}, we define
 \begin{eqnarray}
	\label{eq:taylor,normal}
	\hat{f}(\boldsymbol{r}_j) &=& \boldsymbol{p}_{ij} \cdot 
	\boldsymbol{\delta}f(\boldsymbol{r})|_{\boldsymbol{r} = 
	\boldsymbol{r}_i}, \\
	\label{eq:base,normal}
	\boldsymbol{p}_{ij} &=& (1, r_{1,j} - r_{1,i}, r_{2,j} - r_{2,i},
	\cdots, r_{d,j} - r_{d,i}, (r_{1,j} - r_{1,i})^2, 
	(r_{1,j} - r_{1,i})(r_{2,j} - r_{2,i}), \nonumber\\
	&&\cdots,(r_{1,j} - r_{1,i})^n, (r_{1,j} - r_{1,i})^{n-1}
	(r_{2,j} - r_{2,i}), \cdots, (r_{d,j} - r_{d,i})^n)^{T}, \\
	\label{eq:partial,normal}
	\boldsymbol{\delta} &=& \left(\hat{1},
	\frac{\hat{\partial}}{\hat{\partial} r_{1}},
	\cdots, \frac{\hat{\partial}}{\hat{\partial} r_{d}}, \cdots,
	\frac{1}{n!} \frac{\hat{\partial}^n}{\hat{\partial} r_{1}^n},
	\frac{1}{(n-1)!1!} \frac{\hat{\partial}^n}
	{\hat{\partial} r_{1}^{n-1}\hat{\partial} r_{2}},\cdots,
	\frac{1}{n!}\frac{\hat{\partial}^n}{\hat{\partial} r_{d}^n}\right)
	^{T},
 \end{eqnarray}
where $\hat{1}, \hat{\partial}^m / \hat{\partial} r^m$ are the
approximations of $1, \partial^m / \partial r^m$ and have the error
$\mathcal{O}(\|\boldsymbol{r}\|^{n+1-m})$, and $d$ is the dimension of the
space. Therefore, the right hand side
of equation (\ref{eq:taylor,normal}) matches the Taylor expansion around
$\boldsymbol{r}_j = \boldsymbol{r}_i$ of $f(\boldsymbol{r}_j)$ up to for
($n-m$)-th order for $m$-th order spatial derivatives.

If we do not need the interpolation formula for $f_i$, it is possible
to set $\hat{1} = 1$
(e.g. \cite{tamai2013development}). However, in this paper, we need the
interpolation formula for $f_i$ in section \ref{sec:form_ref}.
Therefore, we do not use this form.

The optimal $\boldsymbol{\delta}f(\boldsymbol{r})|_{\boldsymbol{r} =
\boldsymbol{r}_i}$ for which $\epsilon$ takes the extreme, is given by
 \begin{eqnarray}
	\boldsymbol{\delta}f(\boldsymbol{r})|_{\boldsymbol{r} =
	\boldsymbol{r}_i}
	&=& B^{-1}_i \sum_j \widetilde{W}_{ij} f_j {\boldsymbol{p}}_{ij},\\
	B_i
	&=& \sum_j \widetilde{W}_{ij} {\boldsymbol{p}}_{ij} \otimes
	{\boldsymbol{p}}_{ij},
 \end{eqnarray}
where $B_i$ is a regular matrix. Note that if $B_i$ becomes a rank
deficient, it means that the set of $f_j$ does not have enough information
to derive a unique $\boldsymbol{\delta}f(\boldsymbol{r})|_{\boldsymbol{r} =
\boldsymbol{r}_i}$. Therefore, in such a case, we widen the non-zero region of
$\widetilde{W}_{ij}$ to increase the number of particles in the shape
function.

In this paper, we set $\widetilde{W}_{ij} = W_{ij}$, where $W_{ij}$ is a
kernel function that depends on $\boldsymbol{r}_{ij}$ and $h_i$. Here,
$h_i$ is a kernel length that denotes the width of the kernel function. In
SPH, it is usually given by
\begin{eqnarray}
	h_i = \eta \left(\frac{m_i}{\rho_i}\right)^{1/d},
\end{eqnarray}
where $\eta$ is a constant coefficient. The parameters $m$ and $\rho$ are
the mass and the mass-density.

From the above, the contribution of $f_j$ to $\hat{f}(\boldsymbol{r}_i)$ is
given by
 \begin{eqnarray}
	\label{eq:my,sf0}
	\phi_{ij}
	= \sum_{\alpha} \left[B^{-1}_i\right]_{0 \alpha} {W}_{ij}
	p_{\alpha,ij},
 \end{eqnarray}
and that to derivatives $\partial / \partial \boldsymbol{r}_{\beta}
\left[\hat{f}(\boldsymbol{r}_i)\right]$ is
 \begin{eqnarray}
	\label{eq:my,sf1}
	\psi_{\beta, ij}
	= \sum_{\alpha} \left[B^{-1}_i\right]_{\beta \alpha} {W}_{ij}
	p_{\alpha,ij},
 \end{eqnarray}
where $[B_i]_{\alpha \beta}$ is element $\alpha, \beta$ of matrix
$B_i$ and $p_{\alpha, ij}$ is element $\alpha$ of vector
$\boldsymbol{p}_{ij}$. Note that index $\beta$ takes $1 \leq \beta
\leq d$ in equation (\ref{eq:my,sf1}).

Alternatively, in MLSPH, the derivative of equation (\ref{eq:my,sf0}) is
used as $\partial / \partial \boldsymbol{r}_{\beta} \left[
\hat{f}(\boldsymbol{r}_i)\right]$. In this case, all components of
$\boldsymbol{\delta}f(\boldsymbol{r})|_{\boldsymbol{r} = \boldsymbol{r}_i}$
are freely changed to minimize the residual. Therefore, the error is
smaller than that of equation (\ref{eq:my,sf1}). However, we can get only
zeroth order differential shape function, and thus it takes calculation
cost to
differentiate shape function which consists of a matrix and others.
Therefore, we use equation (\ref{eq:my,sf1}) as a differential shape
function.

In this paper, we define that the scheme is $n$-th order in space, if the
first-order spatial derivative is $n$-th order.

\subsection{Type 2 shape functions}
\label{sec:form_mwr}
We derived the shape functions using MLS in section \ref{sec:form_norm}.
However, the shape functions of CSPM, MSPH and FPM cannot be expressed in
terms of MLS because these functions are derived from the idea different
from MLS. In this section, we derive them using MWR.
In MWR, the weighted approximation error is set zero.
 \begin{eqnarray}
	\label{eq:error,mwr}
	\int \left[f_j - \hat{f}(\boldsymbol{r}_j)\right]
	\widehat{W}_{ij} d\boldsymbol{r}_j = 0.
 \end{eqnarray}
In this section, we show that the shape functions of CPHM, MSPH and FPM can
be derived using the trial function of the form in equation
(\ref{eq:taylor,normal}). Here, since there are $_{(d+n)}C_d$ unknown
parameters, we need to give $_{(d+n)}C_d$ differential forms of
$\widehat{W}_{ij}$. In these methods, these functions are given by
 \begin{eqnarray}
	\boldsymbol{q}_{ij}^t &=&
	(W_{ij}, \nabla_1 W_{ij}, \nabla_2 W_{ij}, \dots, \nabla_d W_{ij},
	\nabla^2_1 W_{ij}, \nabla_1\nabla_2 W_{ij}, \nonumber\\
	&& \dots, \nabla^n_1 W_{ij}, \nabla_1^{n-1}\nabla_2 W_{ij}, \dots,
	\nabla_d^n W_{ij}),
 \end{eqnarray}
Thus, we have
 \begin{eqnarray}
	\int \left[f_j - \hat{f}(\boldsymbol{r}_j)\right]
	\boldsymbol{q}_{ij} d\boldsymbol{r}_j = 0.
 \end{eqnarray}
Note that the $n$-th order differential shape functions in CSPM are derived
recursively by using zero to $(n-1)$ order derivatives. Therefore,
$\hat{f}(\boldsymbol{r}_i)$ is given by setting
$\boldsymbol{q}_{ij}^t = W_{ij}$, and then the first derivatives are
produced by using this $\hat{f}(\boldsymbol{r}_i)$ and setting
$\boldsymbol{q}_{ij}^t = (\nabla_1 W_{ij}, \nabla_2 W_{ij}, \dots,
\nabla_d W_{ij})$.

Finally, the integral $\int f_j d\boldsymbol{r}_j$ is approximated by
summation $\sum_j f_j V_j$, and we have
 \begin{eqnarray}
	\label{eq:sf,mwr}
	\boldsymbol{\delta} f_i &=& B_{i}^{'-1}
	\sum_j f_j V_j \boldsymbol{p}_{ij}, \\
	B'_{i} &=& \sum_j V_j \boldsymbol{p}_{ij} \otimes \boldsymbol{q}_{ij},
 \end{eqnarray}
where $V_j$ is the ``volume'' of particle $j$.
From the above, the contribution of $f_j$ to $\hat{f}(\boldsymbol{r}_i)$ is
given by
 \begin{eqnarray}
	\phi'_{ij} &=& \sum_{\alpha} [B_{i}^{'-1}]_{0\alpha}
	V_j p_{\alpha,ij},
 \end{eqnarray}
and that of derivatives $\partial / \partial \boldsymbol{r}_{\beta}
\hat{f}(\boldsymbol{r}_i)$ is
 \begin{eqnarray}
	\psi'_{\beta,ij} &=& \sum_{\alpha} [B_{i}^{'-1}]_{\beta\alpha}
	V_j p_{\alpha,ij}.
 \end{eqnarray}
In the following, we use the shape functions of equations (\ref{eq:my,sf0})
and (\ref{eq:my,sf1}).

\subsection{The artificial viscosity for multi-dimensions}
\label{sec:form_av}
Many forms of the artificial viscosity have been proposed for SPH to capture
shocks (e.g. \cite{1985MNRAS.215..125L}, \cite{1997JCoPh.136..298M}).
However, most of them cause unwanted shear
viscosity (e.g. \cite{1995JCoPh.121..357B}, \cite{2010MNRAS.408..669C}).

To reduce unwanted shear viscosity, \citet{1995JCoPh.121..357B} and
\citet{2010MNRAS.408..669C} introduced shear switches that reduce viscosity
when the shear exists. Alternatively, \citet{2002JCoPh.179..238I} and
\citet{2015MNRAS.450...53H} proposed the use of Riemann solvers in order
not to use the artificial viscosity. \citet{1989ApJS...70..419H} introduced
a form of the artificial viscosity different
from that of \citet{1983JCoPh..52..374M}. This artificial viscosity 
(hereafter NRAV) was derived by adding the bulk 
viscosity to the artificial viscosity formulated by 
\citet{1950JAP....21..232V}. This viscosity is applied only when
$\boldsymbol{\nabla}\cdot\boldsymbol{v} < 0$, where $\boldsymbol{v}$ is the
velocity. \citet{2016ApJS..224...32H}
compared many different forms of artificial viscosities, including the usual
\citet{1983JCoPh..52..374M} type, and NR type, both with and without
different forms of shear switches and time-dependent switches. They found
the NR type viscosity is the best, when the estimate of
$\boldsymbol{\nabla}\cdot\boldsymbol{v}$ is of high order. When a low-order
estimate is used, it causes unwanted shear viscosity.

These proposed forms of the artificial viscosity can handle fluid with the velocity shear better
than the standard artificial viscosity of SPH can. The standard SPH artificial viscosity is
defined for pairs of particles, and thus there is no easy way to apply it
to high-order schemes in which particles do not have extensive quantities.
Moreover, \citet{2016ApJS..224...32H} demonstrated that even for SPH
schemes, NRAV is better than the standard SPH artificial viscosity. Therefore, we adopted NRAV.
Initially, we used the usual form of NRAV extended to multi-dimensional
space
 \begin{eqnarray}
	\label{eq:avmom,mdim,org}
	\frac{d\boldsymbol{v}}{dt} &=&
	-\frac{1}{\rho}\frac{\partial q^{\rm AV}}{\partial \boldsymbol{r}},\\
	\label{eq:aveng,mdim,org}
	\frac{du}{dt} &=& -\frac{q^{\rm AV}}{\rho}\frac{\partial \boldsymbol{v}}
	{\partial \boldsymbol{r}},\\
	\label{eq:avq,mdim,org}
	q^{\rm AV} &=& -\beta^{\rm AV}{\rho h^2}\frac{\partial \boldsymbol{v}}
	{\partial \boldsymbol{r}}\cdot
	\left|\frac{\partial \boldsymbol{v}}{\partial \boldsymbol{r}}\right|
	\Theta(-{\partial \boldsymbol{v}}/{\partial \boldsymbol{r}}),
 \end{eqnarray}
where $\beta^{\rm AV}$ is a constant coefficient and $u$ is the internal
energy. The function $\Theta(\chi)$ is the Theta function.
We, however, found that this form leads to numerical instability. The
reason why the instability takes place is that the pressure-like quantity
in NRAV in equations (\ref{eq:avmom,mdim,org}) and (\ref{eq:aveng,mdim,org})
is isotropic. Therefore, these ``pressure'' can and does operate to
directions perpendicular to the direction of compression, resulting in the
increase of the kinetic energy.
Therefore, we extend the NRAV to multi-dimensions so that the artificial viscosity operates
only in the direction of the maximum compression. Note that we assume that
the number of shock waves at one position is only one.

Let $X$ the coordinate of the direction of a shock wave.
\citet{1950JAP....21..232V} introduced the following the artificial viscosity
to equations of moment and energy.
 \begin{eqnarray}
	\label{eq:avmom,1dim,org}
	\frac{dv_X}{dt} &=& -\frac{1}{\rho}\frac{\partial q^{\rm AV}}{\partial X},\\
	\label{eq:aveng,1dim,org}
	\frac{du}{dt} &=& -\frac{q^{\rm AV}}{\rho}\frac{\partial v_X}
	{\partial X},\\
	\label{eq:avq,1dim,org}
	q^{\rm AV} &=& -\beta^{\rm AV}{\rho h^2}\frac{\partial v_X}
	{\partial X}\cdot\left|\frac{\partial v_X}{\partial X}\right|
	\Theta(-{\partial v_X}/{\partial X}).
 \end{eqnarray}
To extend this artificial viscosity to multi-dimensions, first, we determine the
direction of $X$ axis along which the fluid is maximally compressed.
Second, the artificial viscosity for this $X$-direction is calculated.
Finally, we transform the calculated the artificial viscosity into the original system of coordinates.
The strain rate tensor is given by
 \begin{eqnarray}
	s_{\alpha \beta} = \frac{1}{2}\left(
	\frac{\partial v_\alpha}{\partial r_\beta}
	+
	\frac{\partial v_\beta}{\partial r_\alpha}
	\right).
 \end{eqnarray}
Using eigenvalues $\lambda_1, \cdots, \lambda_d$ and eigenvectors
$\boldsymbol{b}_1, \cdots, \boldsymbol{b}_d$, $s$ is diagonalized as
\begin{eqnarray}
	\left(
	\begin{array}{cccc}
		\lambda_1 & 0 & \ldots & 0 \\
		0 & \ddots & \ddots & 0 \\
		\vdots & \ddots & \ddots & 0 \\
		0 & \ldots & 0 & \lambda_n
	\end{array}
	\right)
	=
	(\boldsymbol{b}_1, \cdots, \boldsymbol{b}_d)^t s
	(\boldsymbol{b}_1, \cdots, \boldsymbol{b}_d),
\end{eqnarray}
We assume eigenvalues are ordered, $\lambda_1 \leq \lambda_2 \leq \dots \leq
\lambda_d$. Thus, if $\boldsymbol{\nabla}\cdot \boldsymbol{v}< 0$, we apply
the artificial viscosity in the direction of $\boldsymbol{b}_1$. Here, the transformed coordinate
is given by
\begin{eqnarray}
	\boldsymbol{r'} = \boldsymbol{r}(\boldsymbol{b}_1, \boldsymbol{b}_2,
	\dots,\boldsymbol{b}_d).
\end{eqnarray}
First, we derive the artificial viscosity in the equation of motion. In the transformed
coordinate, the artificial viscosity term is given by
\begin{eqnarray}
	\label{eq:av,eom,transformed}
	\frac{d\boldsymbol{v'}}{dt}
	&=& \left(\frac{1}{\rho}\frac{\partial q'}{\partial r'_1},0,\dots,
	0\right),\\
	q'^{\rm AV} &=& -\beta^{\rm AV}{\rho h^2}\lambda_1\left|\lambda_1
	\right|\Theta(-\boldsymbol{\nabla}\cdot\boldsymbol{v}).
\end{eqnarray}
Therefore, in the original coordinate, the artificial viscosity term in the equation of motion is
\begin{eqnarray}
	\frac{d\boldsymbol{v}}{dt}
	= \left(\frac{1}{\rho}\frac{\partial q'}{\partial r'_1},0,\dots,
	0\right)(\boldsymbol{b}_1, \boldsymbol{b}_2,\dots,\boldsymbol{b}_d).
\end{eqnarray}
Here, by using the quantities in the original coordinate,
${\partial q'}/{\partial r'_1}$ can be expressed by
\begin{eqnarray}
	\frac{\partial q'}{\partial r'_1} =
	\frac{\partial q'}{\partial \boldsymbol{r}}\cdot\boldsymbol{b}_1.
\end{eqnarray}
From the above, the artificial viscosity term in the equation of motion in the original
coordinate is given by
 \begin{eqnarray}
	\label{eq:av,mlutdim,org}
	\frac{d\boldsymbol{v}}{dt} = -\boldsymbol{b}'_1\frac{1}{\rho}
	\frac{\partial q'^{\rm AV}}{\partial \boldsymbol{r}}
	\cdot\boldsymbol{b}_1,
 \end{eqnarray}
where $\boldsymbol{b}'_1$ is $(b_{1,1}, b_{2,1}, \dots, b_{n,1})$. Here,
$b_{\alpha, 1}$ denotes the first element of $\boldsymbol{b}_{\alpha}$.
Next, we derive the artificial viscosity in the equation of energy. In this
case, it is the same artificial viscosity
for both the original coordinate and the transformed one because the
equation of energy is scalar. Therefore,
 \begin{eqnarray}
	\frac{du}{dt} = -\frac{q'^{\rm AV}}{\rho}\lambda_1.
 \end{eqnarray}

In this paper, we also use the bulk viscosity introduced by 
\citet{1983JCoPh..52..374M}. Therefore, $q'^{\rm AV}$ changes to
 \begin{eqnarray}
	\label{eq:avGM,mlutdim,org}
	q'^{\rm AV} &=& -\left[\alpha^{\rm AV}\rho c_s h
	+ \beta^{\rm AV} \rho h^2 |\lambda_1|\right]\lambda_1
	\Theta(-\boldsymbol{\nabla}\cdot\boldsymbol{v}),
 \end{eqnarray}
In this paper, we use $\alpha^{\rm AV} = 1$ and $\beta^{\rm AV} = 2$ where
$\alpha^{\rm AV}$ is a constant coefficient, and $c_s$ is the sound velocity.

The introduction of the linear bulk viscosity of equation
(\ref{eq:avGM,mlutdim,org}) implies that the viscosity is active even for
infinitesimal compression. In other words, the fluid is viscous even if
there is no shock. In order to reduce the viscosity in the absence of the
shock, so-called ``bulk switches'' have been proposed
(\cite{1997JCoPh.136...41M}, \cite{2000A&A...360..171R}). We use the switch
$\zeta$ multiplied to $q'^{\rm AV}$. The time evolution of $\zeta$ is given
by
 \begin{eqnarray}
	\frac{d\zeta}{dt} &=& -(\zeta_{\rm max} - \zeta)\max(-\boldsymbol
	{\nabla}\cdot\boldsymbol{v}, 0) - \frac{\zeta - \zeta_{\min}}
	{\tau_{\rm AV}},\\
	\tau_{\rm AV} &=& c_{\tau_{\rm AV}} \frac{h}{c_s},
 \end{eqnarray}
where $c_{\tau_{\rm AV}}$ is a parameter which determines the decay time
scale of the artificial viscosity after shock. In this paper, we use $c_{\tau_{\rm AV}} = 1$ and
$\zeta_{\rm max} = 2$, following \citet{2000A&A...360..171R} and
\citet{2016ApJS..224...32H}. We discuss the choice of $\zeta_{\rm min}$ in
section \ref{sec:test_shock}.

Finally, we introduce a new term which weakens the artificial viscosity when
fluid is compressed in one direction, but is expanding in other directions,
\begin{eqnarray}
	F^{\rm AV} = \left(\frac{|\sum_m \lambda_m|}{\sum_m |\lambda_m|}
	\right)^\upsilon,
 \end{eqnarray}
where $\upsilon$ is a positive coefficient.
Consider the case that $|\lambda_1| = |\lambda_2|$ and $\lambda_1 +
\lambda_2  = 0$. Obviously, $\boldsymbol{\nabla}\cdot\boldsymbol{v} = 0$,
and there is no compression. However, numerically determined
$\boldsymbol{\nabla}\cdot\boldsymbol{v}$ can be negative due to truncation
errors and in that case equation (\ref{eq:avGM,mlutdim,org}) can result in
the strong artificial viscosity. We can reduce this errortic activation of the artificial viscosity using this term.
In this paper, we use $\upsilon = 2$. We multiply $q'^{\rm AV}$ by $\zeta$ and $F^{AV}$.
Thus, in our study, we use
\begin{eqnarray}
	q^{\rm AV} = \zeta F^{\rm AV} q'^{\rm AV},
\end{eqnarray}
instead of $q'^{\rm AV}$.

\subsection{Rearrangement of particles}
\label{sec:form_ref}
As we have discussed in section \ref{sec:intro}, the rearrangement of
particles is necessary to handle large deformations of inviscid fluid,
if we use high-order schemes.
Therefore, in this section, we describe how to rearrange particles. In
this paper, we consider two dimensional cases.

First, we consider periodic boundary. In this case, particles are
rearranged to the initial pattern. Of cause, the initial pattern does not
have large deviations of the number density of particles. For example,
consider the case in which initial pattern is a grid. If the distribution
of particles has become distorted as in the left-hand side panel
in figure \ref{fig:ref,grid}, we rearrange particles in a grid pattern (see
the right in figure \ref{fig:ref,grid}).

\begin{figure}[ht]
	\begin{center}
		\includegraphics[width=8cm,clip]{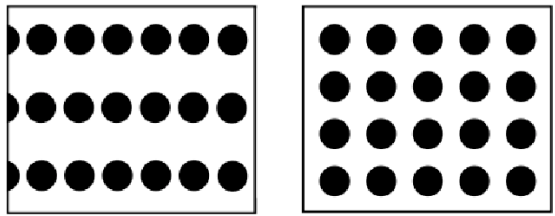} 
	\end{center}
	\caption{An illustration of the rearrangement of particles.
		Particles in a distorted placement (left) is replaced by a
		regular placement (right).}
	\label{fig:ref,grid}
\end{figure}

Next, we consider the fluid with the surface. In this paper, particles
are rearranged to the original grid pattern. Consider the case where particles
are distributed as in the left-hand side panel of figure \ref{fig:ref,surf}. First,
we connect surface particles with lines which are expressed as black lines
in the central panel of figure \ref{fig:ref,surf}. Then, particles are placed on the
intersections of the black line and the grid line, for example, the black dots in the central
panel of figure \ref{fig:ref,surf}. If the interval between particles
is smaller than $0.81 \Delta r_{\rm grid}$, where $\Delta r_{\rm grid}$ is the width
of a grid, one particle is removed. Finally, we put particles on grid points as in
the right-hand side panel of figure \ref{fig:ref,surf}.

\begin{figure}[ht]
	\begin{center}
		\includegraphics[width=12cm,clip]{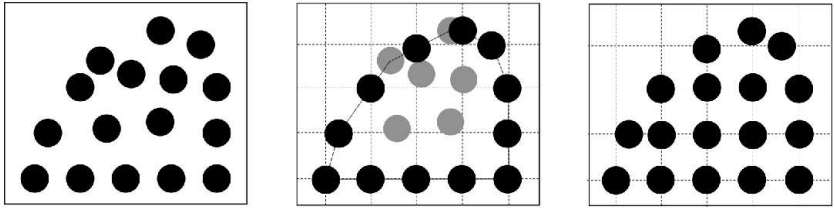}
	\end{center}
	\caption{An illustration of the rearrangement of particles on and
		near the surface. Particles in a distorted placement (left)
		is replaced by a regular placement (right). New particles
		on the surface are generated at the intersection of the
		surface and grid lines.}
	\label{fig:ref,surf}
\end{figure}

The physical quantity, $f_i$, of a new particle $i$ is calculated by using
$f'_j$, where $f'_j$ is the quantity of particle $j$ before rearrangement.
 \begin{eqnarray}
	f_i = \sum_j f'_j \phi_{ij}.
 \end{eqnarray}

Finally, we describe the criterion for rearrangement. Typically,
rearrangement is required when local distribution of particles has become highly
anisotropic. Consider a simple velocity field with linear shear, such as
$v_x = -y$, applied to initial particle distribution of a tilted cartesian grid
(see the left side panel of figure \ref{fig:ref,shear}). Particles move through shear velocity and the
distribution transforms to that given by the central panel of figure
\ref{fig:ref,shear}. Very soon, large difference between distance of
particles in one direction and that in the orthogonal direction develops as shown in
the right panel in figure \ref{fig:ref,shear}.

\begin{figure}[ht]
	\begin{center}
		\includegraphics[width=12cm,clip]{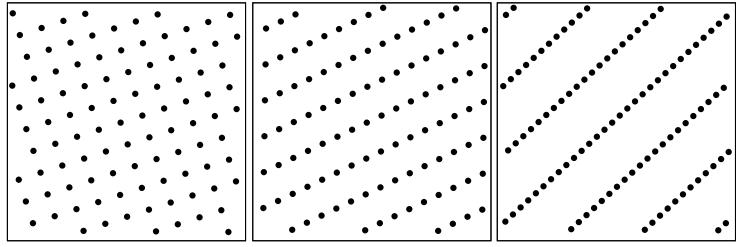} 
	\end{center}
	\caption{The distortion of the distribution  of particles initially
		in the rectangular grid (left) due to the velocity field of
		uniform shear. The center and right panels show the
		distribution of particles after the particles in the top
		and bottom edges move once and twice of the length
		of the edge. }
	\label{fig:ref,shear}
\end{figure}

In order to detect this kind of anisotropy, we use an approximate kernel
weighted moment tensor defined as
 \begin{eqnarray}
	I_i &=& \left(
	\begin{array}{cc}
		\sum_j M_i(x_{ij})^2 W_{ij}
		&
		\sum_j M_i(x_{ij})M_i(y_{ij}) W_{ij}\\
		\sum_j M_i(x_{ij})M_i(y_{ij}) W_{ij}
		&
		\sum_j M_i(y_{ij})^2 W_{ij}
	\end{array}
	\right), \\
	M_i(\chi) &=& {\rm sgn}(\chi)\left(1 - \frac{\chi}{H_i}\right),
 \end{eqnarray}
where ${\rm sgn}(\chi)$ is a sign function, and $x_{ij}$ and $y_{ij}$ are
$(x_j - x_i)$ and $(y_j - y_i)$. The parameter $H_i$ equal to the width of the
kernel function $W_{ij}$. Therefore, if $|\boldsymbol{r}_{ij}|$ is larger
than $H_i$, the value of $W_{ij}$ is zero. The criterion of the
rearrangement is given by 
 \begin{eqnarray}
	\label{eq:rea,criterion}
	\left|1 - \frac{\Lambda_{\rm min}}{\Lambda_{\rm max}} \right| >
	c_{\rm rea},
 \end{eqnarray}
where $\Lambda_{\rm max}$ and $\Lambda_{\rm min}$ are maximum and minimum
eigenvalues of $I_i$, and $c_{\rm rea}$ is a constant coefficient. Note
that it is difficult to apply this criterion to the fluid surface, since
it lacks particles near the surface. Hence,
we apply rearrangement at a constant time interval, when we need to follow
the surface.

\subsection{Fluid equations}
\label{sec:form_sum}
In this section, we derive our high-order mesh-free fully Lagrangian
discretization of equations of continuity, moment and energy. The original set
of partial differential equations are given by
 \begin{eqnarray}
	\label{eq:fluid,orig,eoc}
	\frac{d\rho}{dt}
	&=& -\rho\boldsymbol{\nabla}\cdot\boldsymbol{v},\\
	\label{eq:fluid,orig,eom}
	\frac{d\boldsymbol{v}}{dt}
	&=& -\frac{\boldsymbol{\nabla}P}{\rho},\\
	\label{eq:fluid,orig,eoe}
	\frac{du}{dt}
	&=& -\frac{P}{\rho}\boldsymbol{\nabla}\cdot\boldsymbol{v},
 \end{eqnarray}
where $P$ is pressure.
These equations are discretized using equation (\ref{eq:my,sf1}), and the
artificial viscosity is added. Hence, we have
 \begin{eqnarray}
	\label{eq:fluid_eoc}
	\frac{d\rho_i}{dt}
	&=& -\rho_i\sum_j\boldsymbol{v}_{j}\cdot\boldsymbol{\psi}_{ij},\\
	\label{eq:fluid_eom}
	\frac{d\boldsymbol{v}_{i}}{dt}
	&=& -\frac{1}{\rho_i}\sum_j \left[P_j \boldsymbol{\psi}_{ij}
		+ (q_j^{{\rm AV}}\boldsymbol{\psi}_{ij}
		\cdot\boldsymbol{b}_1)\boldsymbol{b}'_1\right],\\
	\label{eq:fluid_eoe}
	\frac{du_i}{dt} &=&
	-\frac{1}{\rho_i}\sum_j\left(P_i\boldsymbol{v}_{j}\cdot
	\boldsymbol{\psi_{ij}} + q_j^{{\rm AV}}\lambda_1\right).
 \end{eqnarray}
If the fluid is an ideal gas, the equation of state is given by
 \begin{eqnarray}
 	P = (\gamma - 1)\rho u,
	\label{eq:eos_idgas}
 \end{eqnarray}
where $\gamma$ is the ratio of specific heat. Following
\citet{1994JCoPh.110..399M}, we use the equation for the water given by
 \begin{eqnarray}
 	P = C_B\left[\left(\frac{\rho}{\rho_{\rm air}}\right)^7-1\right] +
	P_{\rm air},
	\label{eq:eos_water}
 \end{eqnarray}
where $\rho_{\rm air}$ and $P_{\rm air}$ are density and pressure at
the surface. The constant coefficient $C_B$ is
 \begin{eqnarray}
 	C_B = \frac{200|g|H\rho_{\rm air}}{7}.
  \end{eqnarray}
Here $g$ is the gravitational acceleration and $H$ is the height of a fluid.
For a weakly compressible fluid, we use the linearized equation of state
 \begin{eqnarray}
 	P = c_0^2 (\rho - \rho_{\rm air}),
	\label{eq:eos_weak}
 \end{eqnarray}
where $c_0 = 10\sqrt{|g|H}$.

Following \citet{1989ApJS...70..419H} and \citet{2016ApJS..224...32H}, the
timestep $\Delta t$ is given by
 \begin{eqnarray}
	\label{eq:dt_cfl}
	\Delta t = \min_i C_{\rm CFL}\frac{h_i}{h_i |\boldsymbol{\nabla}\cdot
		\boldsymbol{v}_i| + c_{si} + 1.2\left[\alpha^{\rm AV}c_{si} - \beta^
	{\rm AV}h_i\min(0,\boldsymbol{\nabla}\cdot\boldsymbol{v}_i)
	\right]}.
 \end{eqnarray}
We set the constant coefficient $C_{\rm CFL}$ to $0.3$ unless we state otherwise.

The kernel length $h_i$ is calculated as
 \begin{eqnarray}
	\label{eq:kernellength,cphsf}
	h_i
	&=& \eta\left(\frac{\widetilde{m}_i}{\rho_i}\right)^{1/d},\\
	\label{eq:mass,kernellength,cphsf}
	\widetilde{m}_i
	&=& \rho_{0,i}\prod_{l=1}^d \Delta r_{l,i},
 \end{eqnarray}
where $\Delta \boldsymbol{r}_i$ and $\rho_{0,i}$ are an initial or
rearranged particle spacing and density. In this paper, we set $\eta$ to
$1.6, 2.2, 3.0$ and $3.8$ for first-, second-, third- and fourth-order fitting
formulae in space.

\subsection{Boundary condition}
\label{sec:boundary}
In our method, we can express boundary conditions directly, since our
method is based on fitting polynomials for intensive variables. First, we
consider a free surface. At the free surface, the pressure of the fluid is
same as that of thin air or vacuum. Thus, the boundary condition is given
by
\begin{eqnarray}
	\label{eq:boundary,freesurface}
	P = P_0,
\end{eqnarray}
where $P_0$ is a constant pressure of the assumed air. Now let us consider
a fixed slipping boundary. The boundary condition is
\begin{eqnarray}
	\label{eq:boundary,fixedslipping}
	v_{\perp} = 0.
\end{eqnarray}
Here, $v_{\perp}$ is the velocity perpendicular to the boundary. In the
case of the non-slipping boundary, a fluid element should stay at its
initial point. Therefore, for the non-slip fixed boundary, the boundary
condition is
\begin{eqnarray}
	\boldsymbol{v} = \boldsymbol{0}.
\end{eqnarray}
Finally, at the slipping contact discontinuity, pressure and 
velocity perpendicular to the boundary are continuous. Thus, the boundary
conditions of contact discontinuity of fluid 1 and 2 are
\begin{eqnarray}
	P_1 &=& P_2, \\
	v_{1,\perp} &=& v_{2,\perp},
\end{eqnarray}
where $P_i$ and $v_{i,\perp}$ are pressure and velocity perpendicular
to the discontinuity of the fluid $i$. The boundary conditions at the
non-slipping contact discontinuity are given by
\begin{eqnarray}
	P_1 &=& P_2, \\
	\boldsymbol{v_1} &=& \boldsymbol{v}_2,
\end{eqnarray}
where $\boldsymbol{v}_{i}$ is velocity of the fluid $i$.

\section{Test calculations}
\label{sec:test}
In this section, we show the results of several numerical tests.
First, we compare capabilities of CPHSF and SSPH to handle the fluid
surface by calculating linearize sound modes in section \ref{sec:test_fs}.
In section \ref{sec:test_shock}, the result of the Sod shock tube test is
presented. Here we investigate the errors of the conservation laws.
In section \ref{sec:rotcone}, we show the result of the rotating cone test.
Kelvin-Helmholtz
instability (section \ref{sec:test_khi}) and Rayleigh-Taylor instability
(section \ref{sec:test_rti}) are also calculated. They are suitable to
survey the capability to handle large deformations and fluid instability.
In addition, we investigate if CPHSF can handle free surfaces by gravity
wave test (section \ref{sec:test_gw}) and dam break
test (section \ref{sec:test_dam}). Finally, a cold Keplerian disk is
calculated in section \ref{sec:test_disk}. In these tests, we used the
Backward Euler
integrator for tests with boundary conditions, and we used a third-order
Runge-Kutta method for other tests. For shape functions, we used the
first-order shape function in space unless stated otherwise. The kernel
function is the fourth-order Wendland function
(\cite{wendland1995piecewise}).

\subsection{The one-dimensional behavior of fluid with free surface}
\label{sec:test_fs}
In this section, we investigate how SSPH and CPHSF handle the
fluid surface by analyzing one-dimensional sound wave modes. We assume the
fluid is water with equation of state given by equation
(\ref{eq:eos_water}) with $P_{\rm air} = 0$ and $\rho_{\rm air} = 1000$. The
position, velocity, density and pressure of particle $i$ in the equilibrium
are given by $x_i$, $v_0$, $\rho_0$ and $P_0$. In addition, perturbations
are $\delta x_i$, $\delta v_i$, $\delta \rho_i$ and $\delta P_i$.
The computational domain is $0 \leq x \leq 1$. The number of particles is 101.
The values of $v_0$, $\rho_0$ and $P_0$ are $0$, $1000$ and $0$ for all $i$.

\subsubsection{Standard SPH}
\label{sec:test_fs_ssph}
\begin{figure}[ht]
	\begin{center}
	\includegraphics[width=6cm,clip]{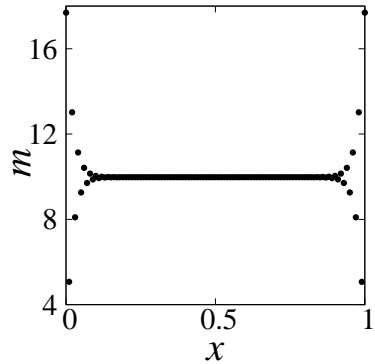}
	\end{center}
	\caption{Mass distribution which satisfies equation (\ref{eq:unit,ssph})
		with the equal particle spacing.
		}
	\label{fig:fs_ssphmass}
\end{figure}

First, we derive the linearized equation of the density perturbation in SSPH.
The fluid equation of SSPH is given by
 \begin{eqnarray}
	\label{eq:x,ssph}
	\frac{dx_i}{dt} &=& v_i, \\
	\label{eq:dens,ssph}
	\rho_i &=& \sum_j m_j W_{ij}, \\
	\label{eq:acc,ssph}
	\frac{dv_i}{dt} &=& -\sum_j{m_j}\left(\frac{P_j}{\rho_j^2} +
	\frac{P_i}{\rho_i^2}\right) \nabla W_{ij}.
 \end{eqnarray}
Note that the value of $\rho_i$ given by equation (\ref{eq:dens,ssph}) does
not become $1000$ because the approximation has zeroth-order error. Therefore,
we derive $m_i$, which satisfies $\rho_i = 1000$ for all $i$, by implicitly
calculating the following equation,
 \begin{eqnarray}
	\label{eq:unit,ssph}
	1000 = \sum_j m_j W_{ij}.
 \end{eqnarray}
Alternatively, we could adjust the locations of particles so that equation
(\ref{eq:unit,ssph}) is satisfied for equal-mass particles, but we chose to
change mass for simplicity. In this test, mass distribution becomes as
figure \ref{fig:fs_ssphmass}.

The perturbation equation for equations (\ref{eq:x,ssph}) to
(\ref{eq:acc,ssph}) is given by
 \begin{eqnarray}
	\frac{d^2\delta {\rho}_i}{dt^2} 
	= \frac{c_{s0}^2}{\rho_0^2} \sum_j\sum_k m_jm_k \left(-\nabla W_{jk}
	(\nabla W_{ik} - \nabla W_{ij})\delta {\rho}_j - \nabla W_{ik} 
	\nabla W_{ij}(\delta {\rho}_j + \delta {\rho}_i)\vphantom{\sum}
	\right), \nonumber\\
 \end{eqnarray}
 where $c_{s0}$ is the sound velocity of the equilibrium state. 

\subsubsection{CPHSF}
\label{sec:test_fs_cphsf}
We derive the linearized equation of the density perturbation in CPHSF.
The fluid equation of CPHSF is given by equations (\ref{eq:fluid_eoc}),
(\ref{eq:fluid_eom}) and (\ref{eq:fluid_eoe}). The linearized perturbation
equation is given by
\begin{eqnarray}
	\frac{d^2\delta \rho_i}{dt^2} 
	= c_{s0}^2 \sum_j\sum_k \delta \rho_k \psi_{jk} \psi_{ij}.
\end{eqnarray}
Equation (\ref{eq:boundary,freesurface}) is used for the boundary
condition. Therefore the value of $\delta \rho_i$ is set zero at $x_i = 0$,
$L$, where $L$ is the width of the fluid and equal $1$ in this test.

We used CPHSF of first- and third-order in space.

\subsubsection{Exact solution}
\label{sec:test_fs_ana}
The perturbation equation at the continuous limit is the wave equation,
\begin{eqnarray}
	\frac{d^2\delta \rho}{dt^2} 
	&=& c_{s0}^2 \nabla^2 \delta \rho.
\end{eqnarray}
The boundary condition is $\delta \rho = 0$ at $x = 0$ and $x = L$ as that
of CPHSF. Consequently, the general solution is given by
\begin{eqnarray}
	\delta \rho &=& \sum_k A(k)\sin\left(\frac{2\pi k}{L}x\right)e^{\omega t} \ \ k \in \mathbb{N},\\
	\omega &=& ic_{s0} \frac{2\pi k}{L},
\end{eqnarray}
where $A(k)$ is the amplitude of mode with wave number $k$.

\subsubsection{Results}
\label{sec:test_fs_res}
\begin{figure}[ht]
	\begin{center}
	\includegraphics[width=6cm,clip]{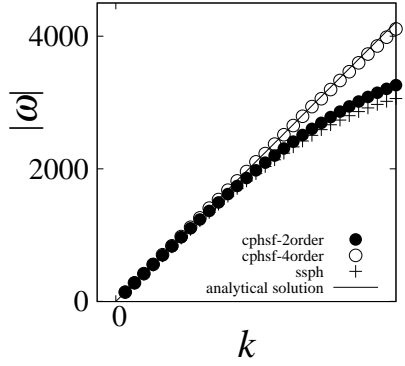}
	\end{center}
	\caption{The absolute value of the frequency $\omega$ plotted against
		the number of the wave $k$ for SSPH and first and third
		space-order CPHSF.}
	\label{fig:fsana_vl}
\end{figure}

\begin{figure}[ht]
	\begin{center}
	\includegraphics[width=12cm,clip]{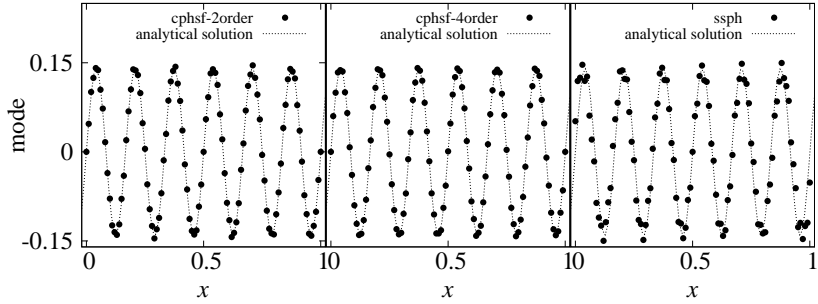} 
	\end{center}
	\caption{The eigenfunction for $k = 6$. The left and central
		panels are for CPHSF with first- and third-order in space.
		The right is for SSPH.}
	\label{fig:fsana_vc}
\end{figure}

\begin{figure}[ht]
	\begin{center}
	\includegraphics[width=12cm,clip]{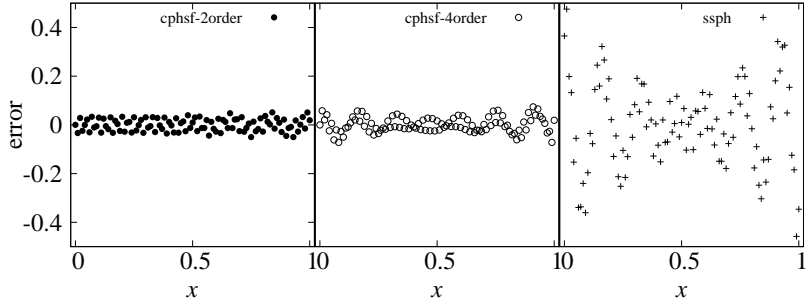} 
	\end{center}
	\caption{The same as figure \ref{fig:fsana_vc} but show the errors
		of eigenfunctions.}
	\label{fig:fsana_vc_error}
\end{figure}

Figure \ref{fig:fsana_vl} shows the eigenvalues of the modes as function of
wave number $k$. We can see that the third-order CPHSF gives very accurate angular
frequency even for large wave numbers. The first-order CPHSF and standard SPH give
similar errors.
Figure \ref{fig:fsana_vc} shows the eigenfunctions of $k = 6$, and
figure \ref{fig:fsana_vc_error} gives the error of eigenfunctions. From these figures,
we can see that the error of eigenfunction calculated with CPHSF is much smaller than
that with SSPH, even when the spatial order of CPHSF is one.

We can conclude that CPHSF is more accurate than SSPH, in particular near the boundary,
even when the spatial order is low.

\subsection{Sod shock tube test}
\label{sec:test_shock}
In this section, we present the result of the Sod shock tube test (\cite{sod1978survey}).
First, we determine the value of $\zeta_{\rm min}$ so that after-shock
numerical oscillation is suppressed. Second, we investigate the errors of
conserved quantities. We assume that fluid is ideal gas with $\gamma = 1.4$.

The computational domain
is $-0.5 \leq x < 0.5$ with a periodic boundary condition, and the
initial boundary of two fluids is at $x = -0.5, 0$. In this test, we use
equal-mass particles. The numbers of particles are $800$ (high density region)
and 200 (low density region), and the total number of
particles is $1000$. Initial velocity is given by $v_x = 0$. The density
is smoothed by a polynomial, and it is given by
 \begin{eqnarray}
	\label{eq:dens,shock}
	\rho(x) = \left\{
	\begin{array}{ll}
		\rho_h,  & -0.5 + x_0 \leq x < -x_0 \\
		\rho_l, & x_0 \leq x < 0.5 - x_0 \\
		\frac{\rho_h - \rho_l}{4}\left[\left(\frac{x}{x_0}\right)^3
		- \frac{3x}{x_0}\right] + \frac{\rho_h + \rho_l}{2}, 
		& {\rm otherwise}
	\end{array}
	\right.
 \end{eqnarray}
where $\rho_h$ and $\rho_l$ are the initial density of the high- and
low-density regions. We used $\rho_h = 1$ and $\rho_l = 0.25$. The parameter
$x_0$ represents the width of the smoothing region, and the value is given
by $x_0 = 0.6(h_h + h_l)$, where $h_h$ and $h_l$ are the
kernel length in the high- and low-density regions.
The position of particle $i$ in the smoothing region is determined so that
they satisfy
\begin{eqnarray}
	\int^{x_i}_{x_{i-1}} \rho(x) dx = \widetilde{m}_i.
\end{eqnarray}
The smoothed pressure is given by
 \begin{eqnarray}
	\label{eq:pres,shock}
	P(x) = \left\{
	\begin{array}{ll}
		P_h, &-0.5 + x_0 \leq x < -x_0 \\
		P_l, & x_0 \leq x < 0.5 - x_0 \\
		\frac{P_h - P_l}{4}\left[\left(\frac{x}{x_0}\right)^3
		- \frac{3x}{x_0}\right] + \frac{P_h + P_l}{2},
		& {\rm otherwise}\\
	\end{array}
	\right.
\end{eqnarray}
where $P_h$ and $P_l$ are the initial pressure of the high- and low-density
regions. We used $P_h = 1$ and $P_l = 0.1795$.

\begin{figure}[ht]
 \begin{minipage}{0.5\hsize}
	\begin{center}
	\includegraphics[width=8cm,clip]{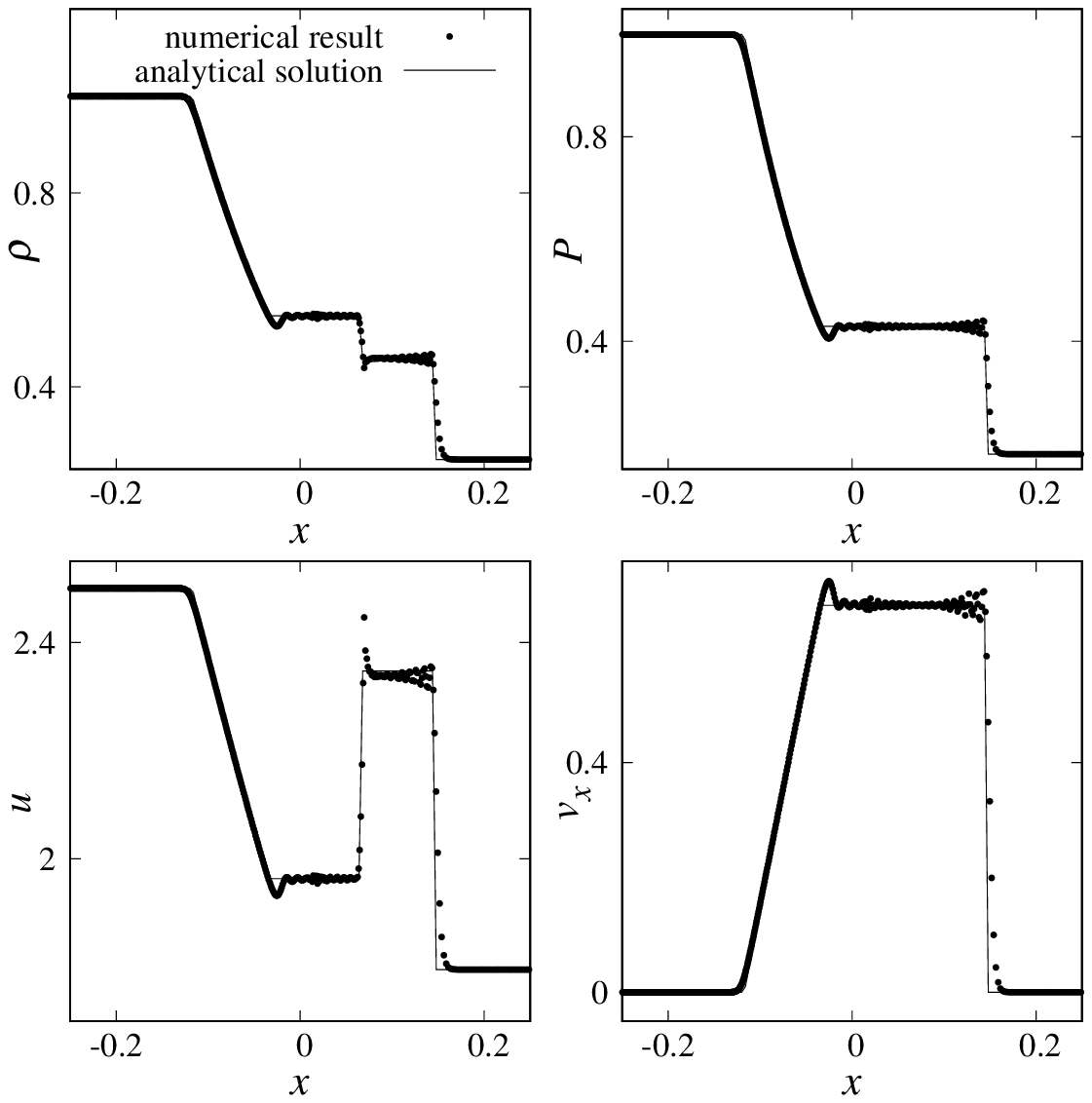} 
	\end{center}
	\caption{Result of the Sod shock tube test at $t=0.1$ with
		first-order CPHSF. We used  $\zeta_{\rm min} = 0.1$.}
	\label{fig:sodshock101av}
 \end{minipage}
 \begin{minipage}{0.5\hsize}
	\begin{center}
	\includegraphics[width=8cm,clip]{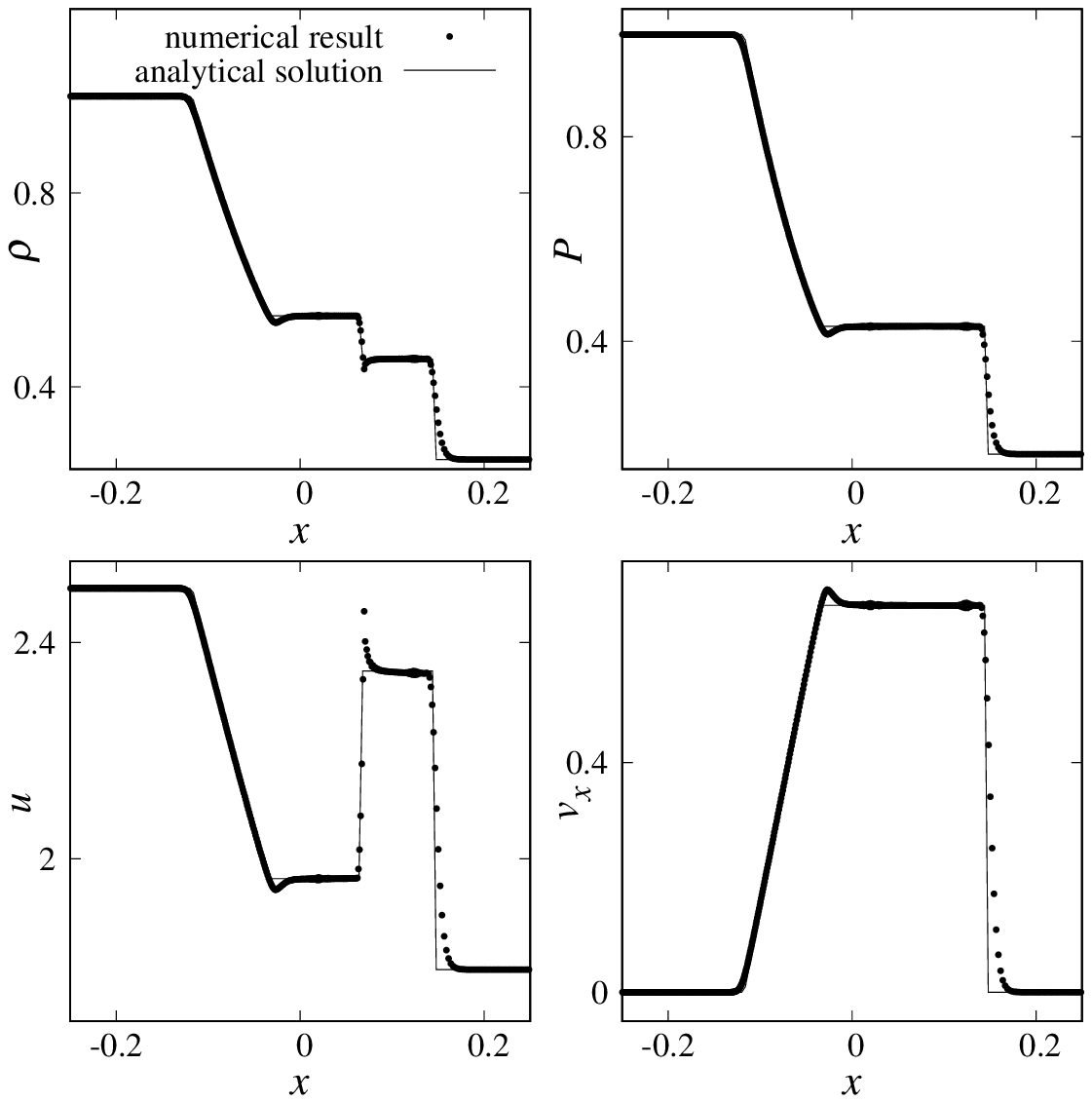} 
	\end{center}
	\caption{The same as figure \ref{fig:sodshock101av}, but for
		$\zeta_{\rm min} = 0.5$.}
	\label{fig:sodshock105av}
 \end{minipage}
\end{figure}

Figure \ref{fig:sodshock101av} shows the numerical solution at $t = 0.1$
with first-order CPHSF with $\zeta_{\rm min} = 0.1$. The post shock oscillation is rather
strong. Figure \ref{fig:sodshock105av} shows the numerical solution for
$\zeta_{\rm min} = 0.5$. In this figure, the post shock oscillation is
suppressed. We use $\zeta_{\rm min} = 0.5$ for other tests in this paper
unless stated otherwise.

We now investigate the errors in conserved quantities. We used first- and
third-order schemes. We set $\zeta = 1$ in these
tests. In the following, we report
the result of two series of test calculations. We varied the number of
particles from $N=1000$ to $8000$
for one test and from $N=1000$ to $16000$ for the other. The kernel length
is calculated using equations (\ref{eq:kernellength,cphsf}) and
(\ref{eq:mass,kernellength,cphsf}). In the first series, we used the NRAV
of the form described in section \ref{sec:form_av}. This means that the
strength of AV is weaker for higher resolution, so that the number of
particles used to resolve shock is approximately constant. In the second
series, we fixed the value of $h$ for equation (\ref{eq:avGM,mlutdim,org})
and initial smoothing parameter $x_0$ in equations (\ref{eq:dens,shock})
and (\ref{eq:pres,shock}),
so that the physical thickness of the shock is independent of $N$. We
used the second series to test the convergence of our scheme.

\begin{figure}[ht]
 \begin{minipage}{0.5\hsize}
	\begin{center}
	\includegraphics[width=8cm,clip]{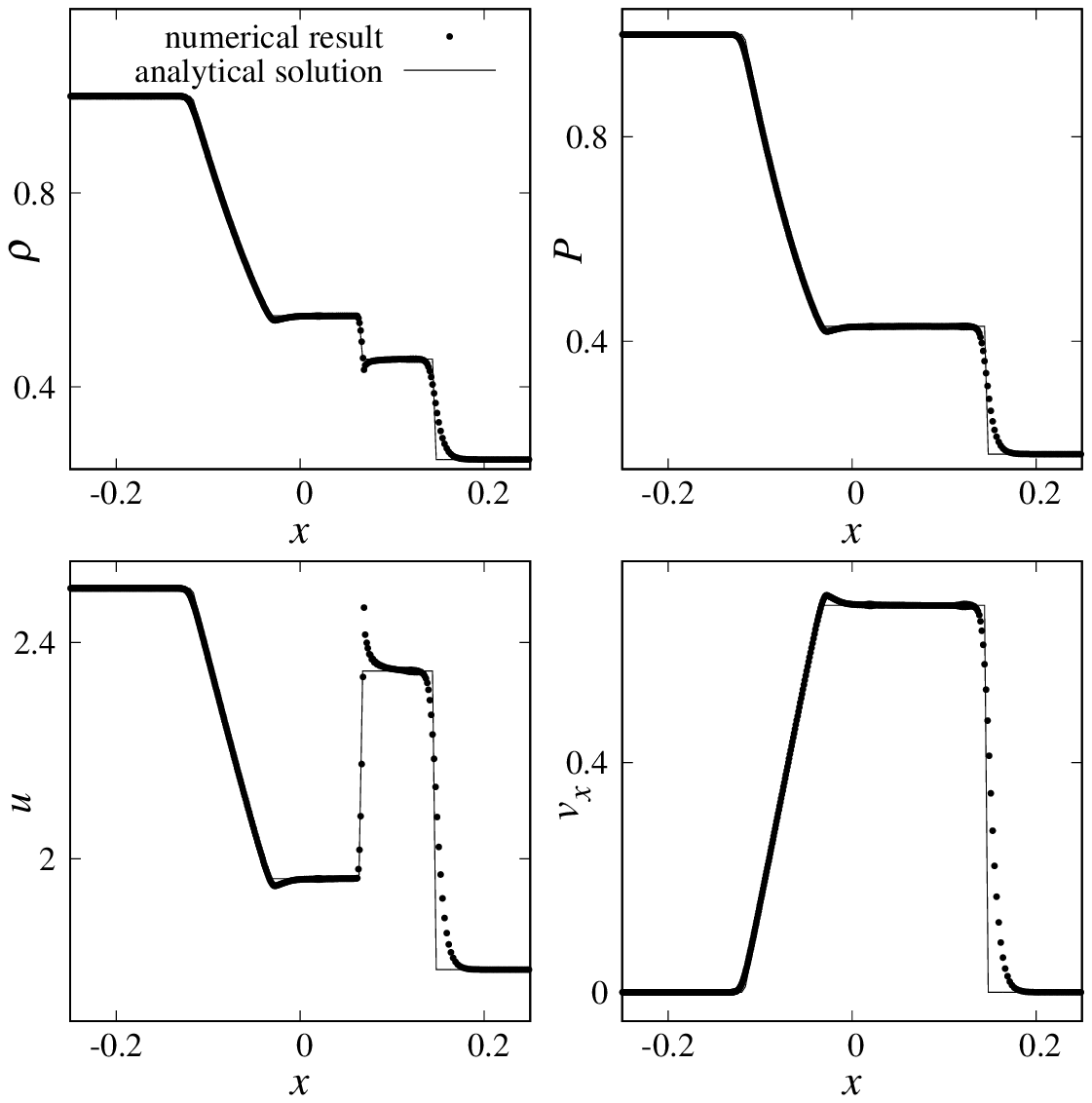}
	\end{center}
	\caption{Result of test 1 at $t=0.1$ with first-order CPHSF.}
	\label{fig:sodshock1s1N_valh}
 \end{minipage}
 \begin{minipage}{0.5\hsize}
	\begin{center}
	\includegraphics[width=8cm,clip]{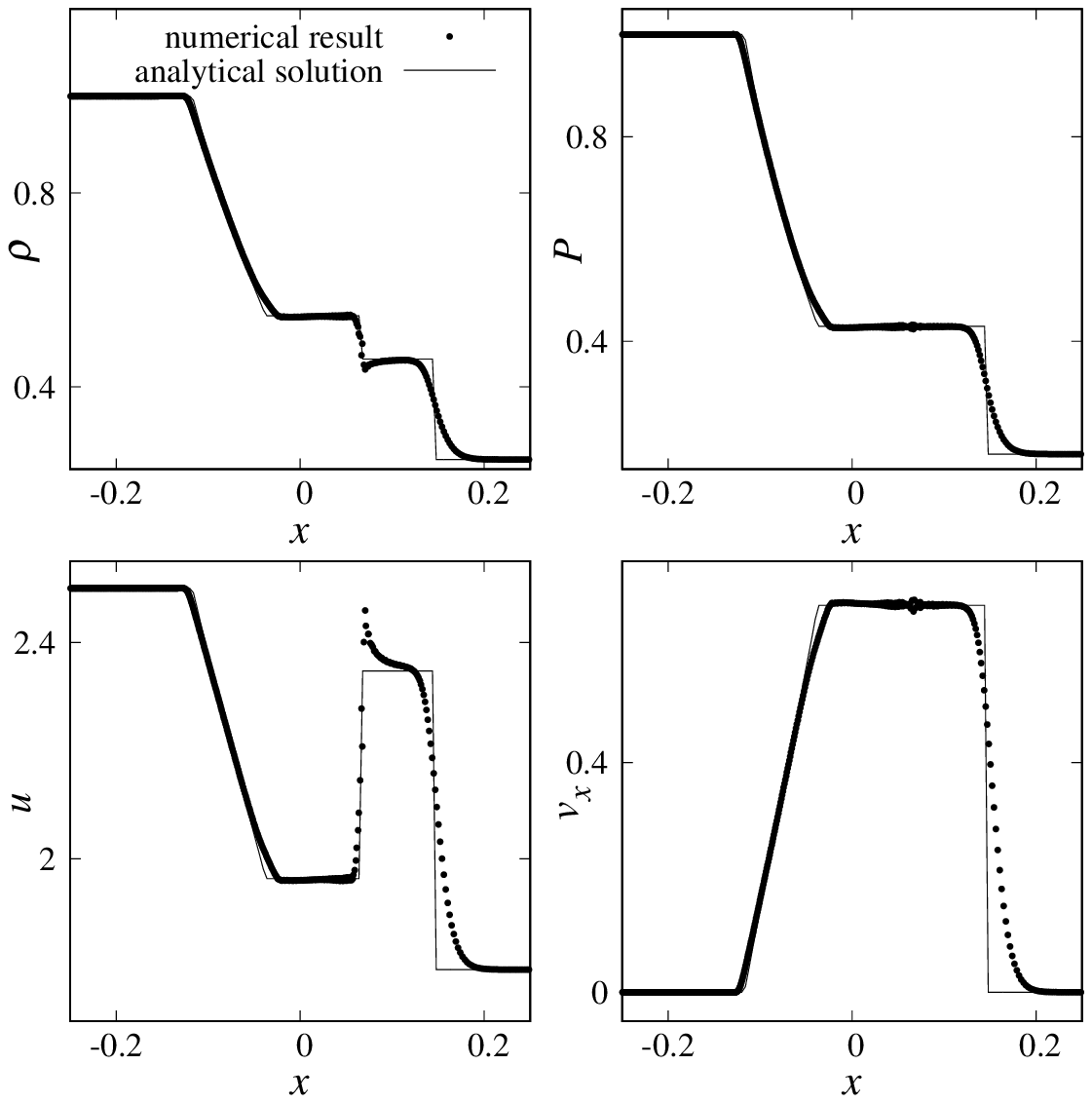}
	\end{center}
	\caption{Same as figure \ref{fig:sodshock1s1N_valh}, but
		the results with third-order CPHSF.}
	\label{fig:sodshock3s1N_valh}
 \end{minipage}
\end{figure}

\begin{figure}[ht]
 \begin{minipage}{0.5\hsize}
	\begin{center}
	\includegraphics[width=8cm,clip]{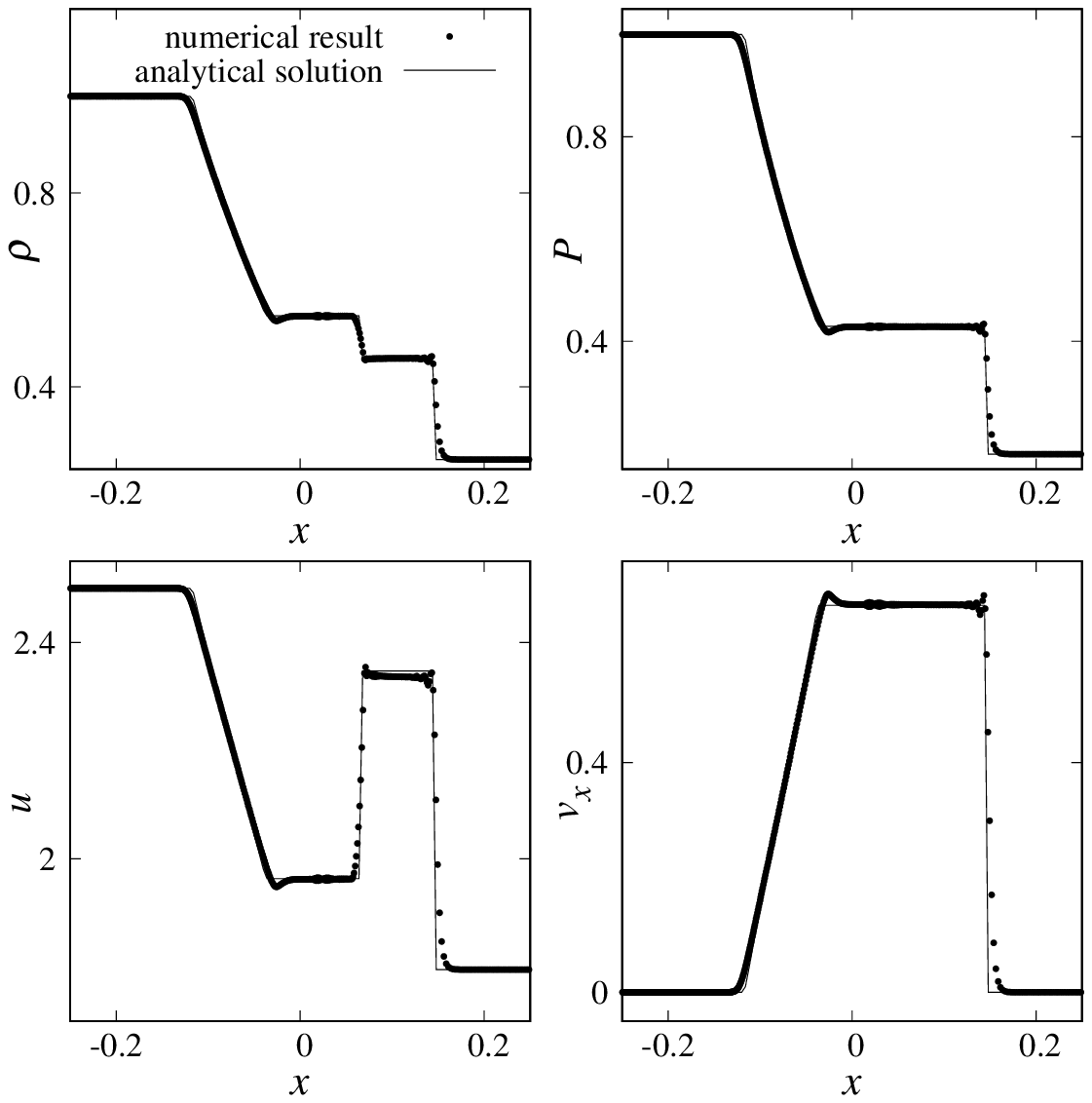}
	\end{center}
	\caption{Result of test 2 at $t=0.1$ with first-order CPHSF.}
	\label{fig:sodshock1s1N_conh}
 \end{minipage}
 \begin{minipage}{0.5\hsize}
	\begin{center}
	\includegraphics[width=8cm,clip]{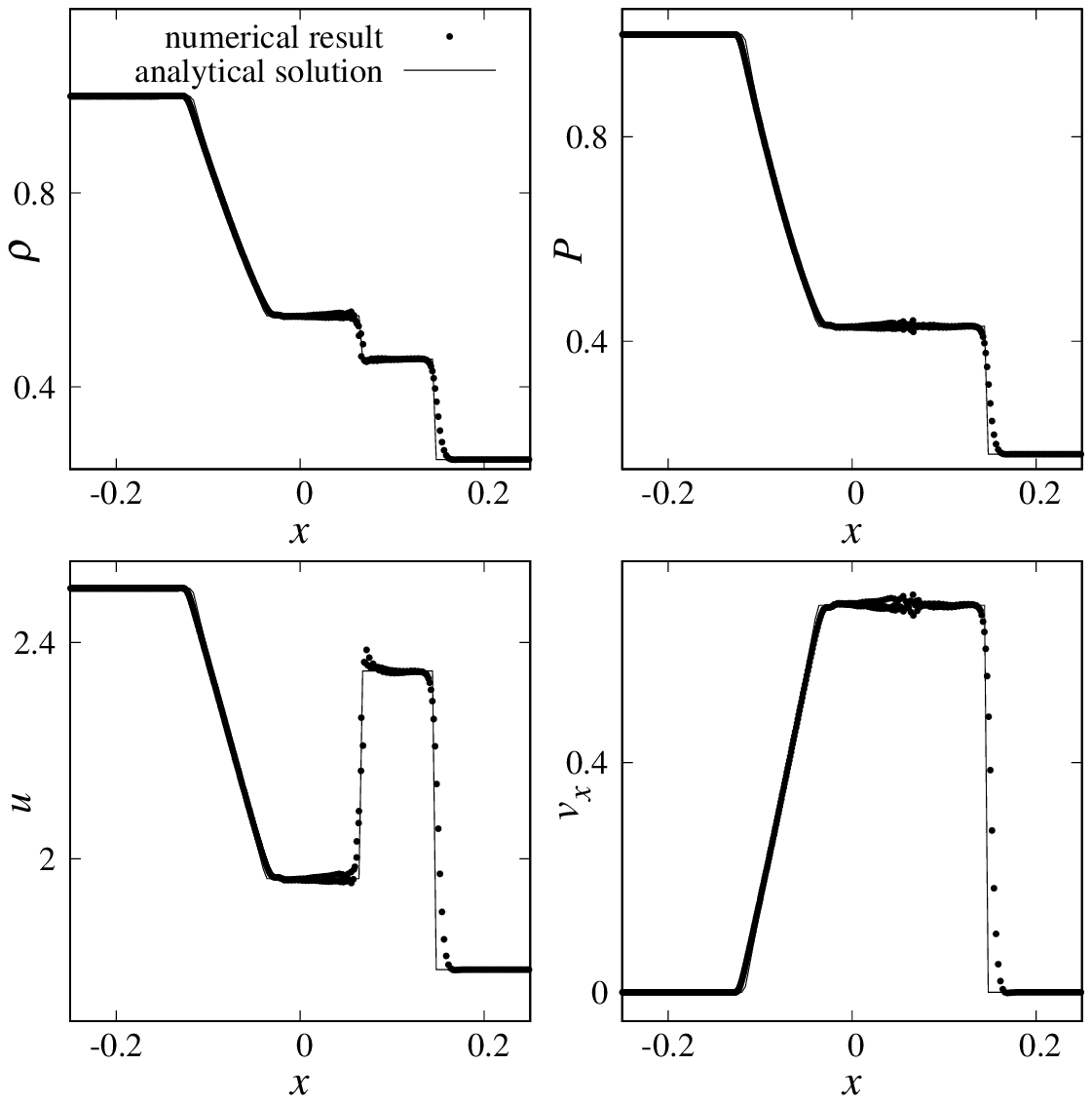}
	\end{center}
	\caption{Same as figure \ref{fig:sodshock1s1N_conh}, but
		the results with third-order CPHSF.}
	\label{fig:sodshock3s1N_conh}
 \end{minipage}
\end{figure}

Figures \ref{fig:sodshock1s1N_valh} to \ref{fig:sodshock3s1N_conh} give the
calculation result for $N=1000$ and $t=0.1$, for tests 1 and 2 and first- and
third- order schemes. We can see that all four results are good in capturing
shocks. The shock is broader for test 1 than for test 2, since the
coefficient for the artificial viscosity is larger for test 1. On the other
hand, weak oscillation is visible in the left-hand-side region of the
contact discontinuity, in particular for the third-order scheme.

In test 1, we compare results with $N = 1000$, $2000$, $4000$, $8000$ to the
analytical solution. Since it is difficult to derive the analytical
solution for test 2, we compare the results with $N = 1000$, $2000$, $4000$, $8000$,
$16000$ to that with $N = 32000$.
We calculate the errors of the total energy and the total momentum in the region
$-0.25 \leq x \leq 0.25$ as the measure of the conservative quantities.
These errors are given by
 \begin{eqnarray}
	\label{eq:error,ene}
	\epsilon_{\rm ene} &=& \frac{|E - E_{a}|}{|E_{0.1}|},\\
	\label{eq:error,mom}
	\epsilon_{\rm mom} &=& \frac{|p - p_{a}|}{|p_{0.1}|},
 \end{eqnarray}
where $E$ and $p$ are numerical solutions of energy and the momentum, and
$E_{a}$ and $p_{a}$ are analytical solutions of energy and the momentum (test 1)
or the results of $N = 32000$ calculation (test 2).
The parameters $E_{0.1}$ and $p_{0.1}$ are the values of $E_{a}$ and
$p_{a}$ at $t = 0.1$. Energy and the momentum of the numerical solution are calculated by 
\begin{eqnarray}
	 E &=& \int^{0.25}_{-0.25} \hat{u} + \frac{1}{2}\hat{\rho}\hat{v}_x^2 dx, \\
	 p &=& \int^{0.25}_{-0.25} \hat{\rho}\hat{v}_x dx,
 \end{eqnarray}
where $\hat{u}$, $\hat{\rho}$ and $\hat{v}_x$ are internal energy, density
and velocity of the numerical solution. Using the shape functions, this integral
is expressed as
 \begin{eqnarray}
	E &\simeq& \sum_m^n \sum_{i, |x_i| \leq 0.25}
	 \frac{\left[\left({x_{i+1} - x_i}\right)^{m+1} - 
	 \left({x_{i} - x_{i-1}}\right)^{m+1}\right]}{2^{m+1}(m+1)}
	\sum_j \left(u_j + \frac{\rho_j}{2}v_{x,j}^2\right) \Psi_{m,ij},\\
	p &\simeq& \sum_m^n \sum_{i, |x_i| \leq 0.25}
	 \frac{\left[\left({x_{i+1} - x_i}\right)^{m+1} - 
	 \left({x_{i} - x_{i-1}}\right)^{m+1}\right]}{2^{m+1}(m+1)}
	\sum_j \rho_j v_{x,j} \Psi_{m,ij},
 \end{eqnarray}
where $n$ is the spatial order of the scheme. The parameter $\Psi_{m,ij}$
is the $m$-th differential-order shape function.

\begin{figure}[ht]
	\begin{center}
	\includegraphics[width=12cm,clip]{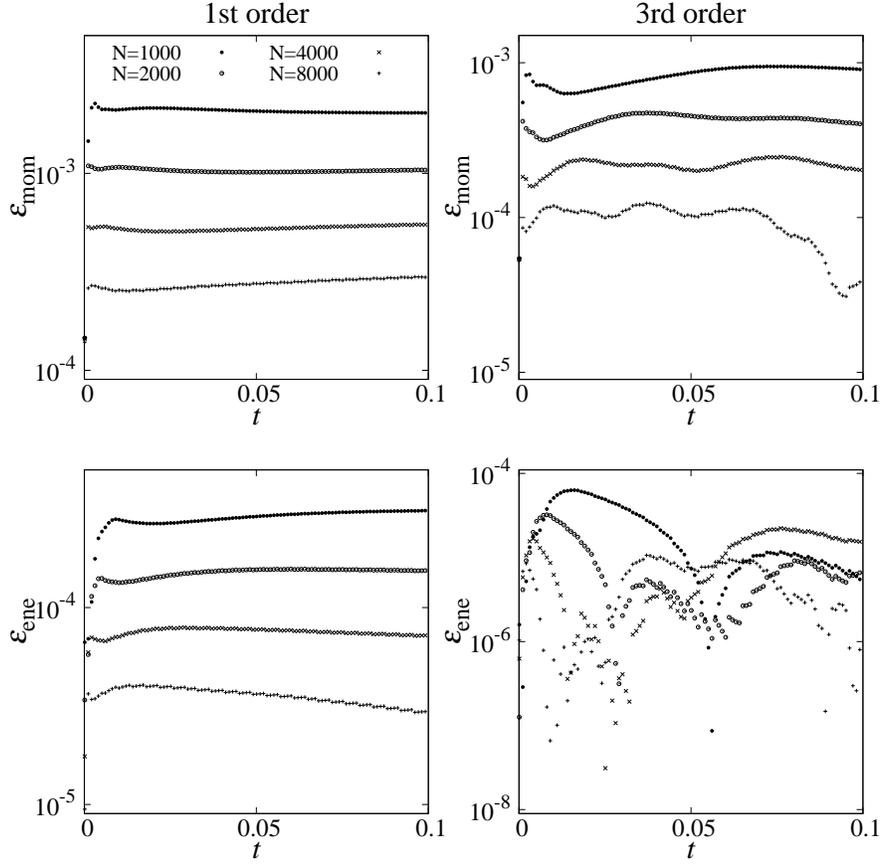} 
	\end{center}
	\caption{
		The upper panels show $\epsilon_{\rm mom}$ and the lower
		panels show $\epsilon_{\rm ene}$. The left-hand side is the
		result of first-order CPHSF and the right is that of
		third-order one.
		}
		\label{fig:shockconsvalh}
 \end{figure}

Figure \ref{fig:shockconsvalh} shows the time evolutions of $\epsilon_{\rm ene}$
and $\epsilon_{\rm mom}$ with first- and third-order schemes and $N = 1000$,
$2000$, $4000$, $8000$ for test 1. We can see that the error is proportional to $1/N$.
In test 1, the initial smoothing length $x_0$ and the strength of the
artificial viscosity depend on $1/N$ in a linear way, and thus the errors
of the fluid equation and artificial viscosity are of zeroth order. The width of
region, where these error occurs, is proportional to $1/N$.
Therefore, the total errors at the shock become $\mathcal{O}(1/N)$.

\begin{figure}[ht]
	\begin{center}
	\includegraphics[width=12cm,clip]{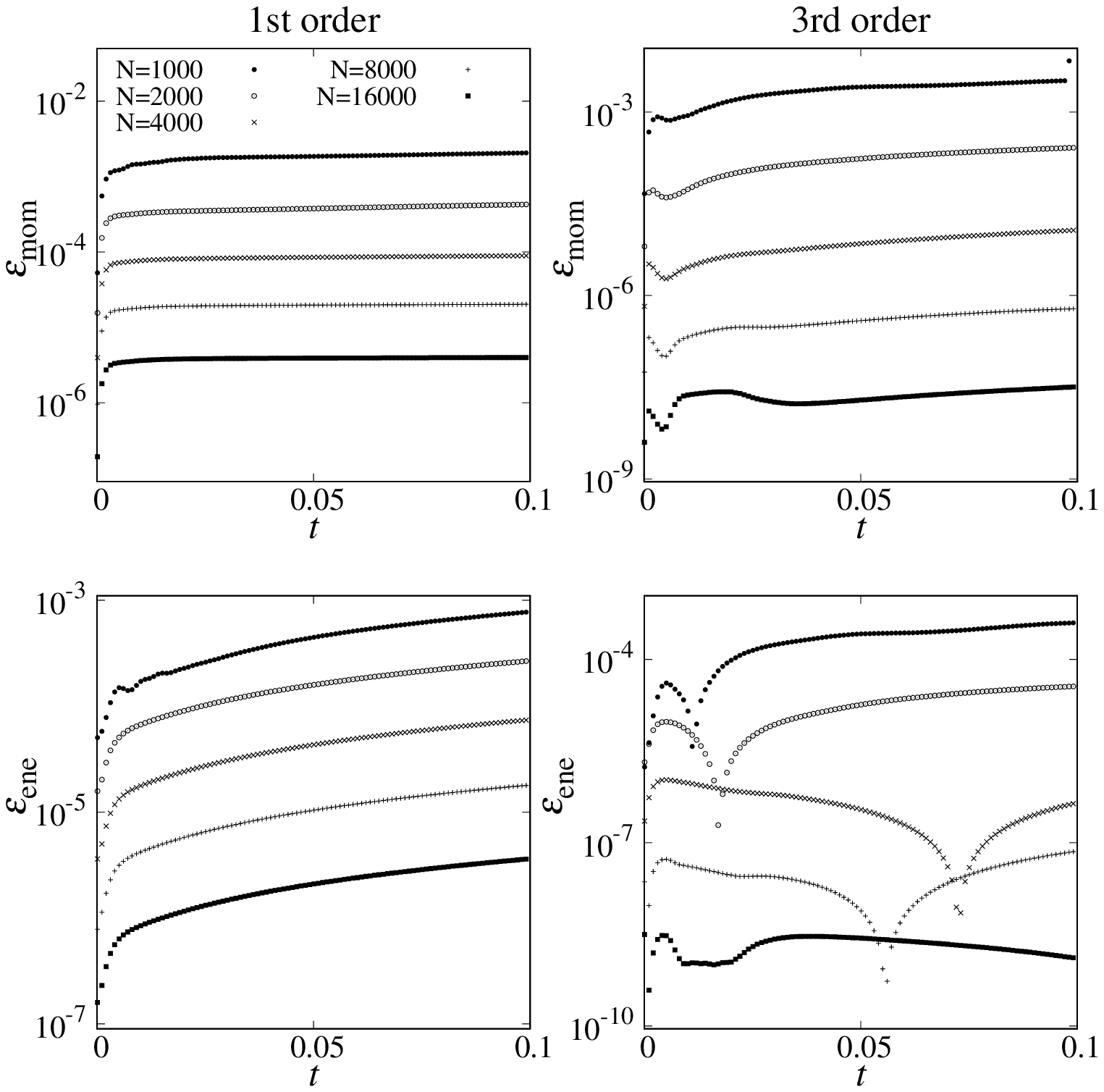} 
	\end{center}
	\caption{Same as figure \ref{fig:shockconsvalh}, but for test 2.}
		\label{fig:shockconsconh}
 \end{figure}

Figure \ref{fig:shockconsconh} show the time evolutions of $\epsilon_{\rm ene}$
and $\epsilon_{\rm mom}$ with first- and third-order schemes and $N = 1000$,
$2000$, $4000$, $8000$ $16000$ for test 2. We can see that the error is much
smaller than those in figure \ref{fig:shockconsvalh}. If the result converge to an 
exact solution following the order of the scheme, the error relation to
$N=32000$ result should be given by $\epsilon \propto 1/N^{n+1} - 1/(32000)^{n+1}$
where $n$ is the order of the scheme.

\begin{figure}[ht]
	\begin{center}
	\includegraphics[width=12cm,clip]{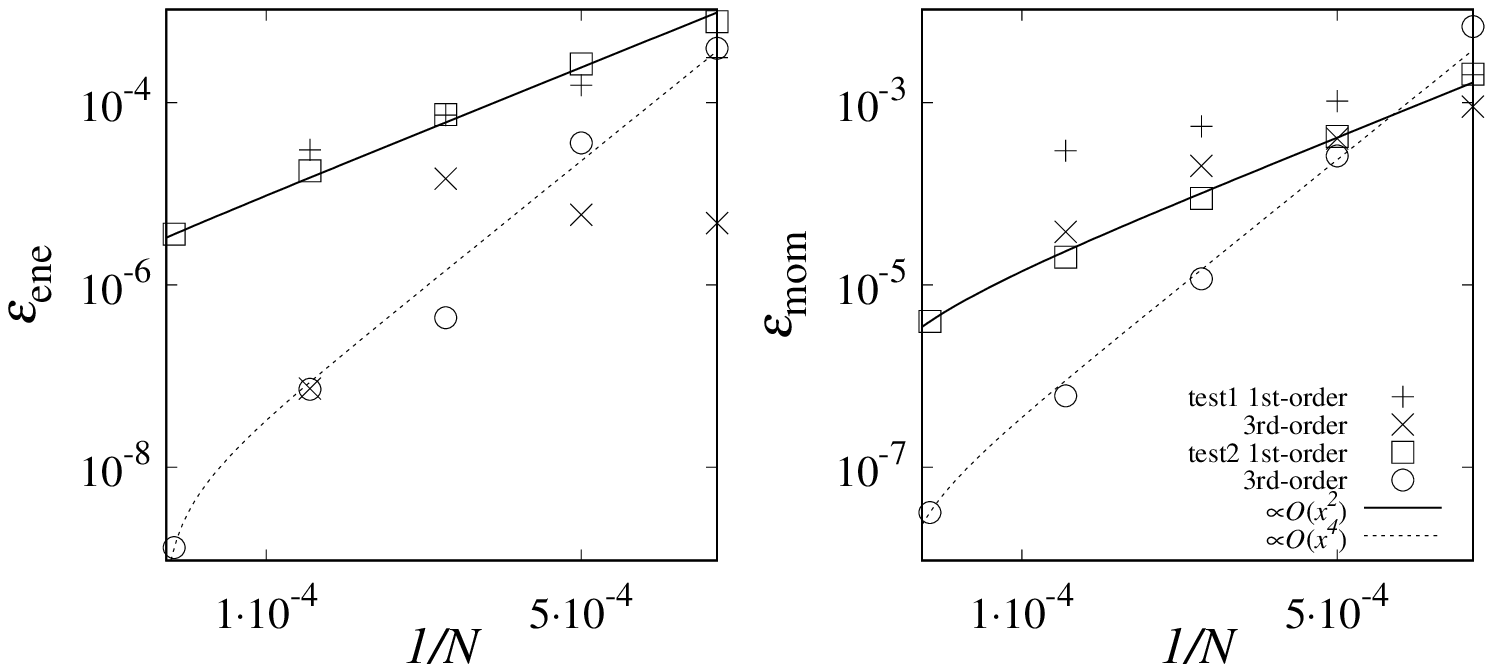} 
	\end{center}
	\caption{
		The errors $\epsilon_{\rm ene}$ (left) and
		$\epsilon_{\rm mom}$ (right) at $t=0.1$ plotted against
		$1/N$. Crosses and ``X''s show the results with first- and
		third-order CPSHSF for test 1, and squares and circles
		those for test 2. Solid and dashed curves show the
		theoretical models for the error of first- and third-order
		schemes for test 2.
		}
		\label{fig:shockconsNN}
 \end{figure}

In figure \ref{fig:shockconsNN}, we plot
the error at $t = 0.1$ as a function of $N$. We can see that the result of
test 2 with third-order scheme actually shows the $\mathcal{O}(N^{-4})$
error, demonstrating that the spatial order of our CPHSF scheme is
consistent with the numerical order with of the fitting polynomial.

\subsection{Rotating cone test}
\label{sec:rotcone}
In this section, we present the result of  the rotating cone test (e.g. 
\cite{crowley1968numerical}, \cite{chock1991comparison}, \cite{vijay1998fluid})
to discuss
the effect of the numerical diffusion caused by the rearrangement.
In this test, the rigid rotation a cone-shaped object is followed for
two complete circles. Since CPHSF is fully Lagrangian scheme, without
the rearrangement of particles it will perfectly conserve the initial
shape of the object. We forced the periodic rearrangement of particles
to see its effect.

We used the initial condition the same as in \citet{vijay1998fluid}.
The computational domain is $-32 \leq x \leq 32, -32 \leq y \leq 32$.
The initial density distribution is given by
\begin{eqnarray}
	\rho(r_c) &=&
	\left\{
	\begin{array}{ll}
		\frac{\rho_{\rm back} - \rho_{\rm peak}}{\Delta r}
		r_c + \rho_{\rm peak} & r_c < \Delta r \\
		\rho_{\rm back} & {\rm otherwise}
	\end{array}
	\right. \\
	r_c &=& \sqrt{(x-16)^2 + y^2},
\end{eqnarray}
where $\rho_{\rm peak} = 100$, $\rho_{\rm back} = 5$ and
$\Delta r = 32 / (\sqrt{N} - 1)$ where $N$ is the number of particles.
Initial velocity is
the rigid body rotation given by
\begin{eqnarray}
	v_x &=& -\omega_{\rm cone}y, \\
	v_y &=& \omega_{\rm cone}x,
\end{eqnarray}
where $\omega_{\rm cone}$ is the angular velocity and we set $\omega_{\rm cone}
= 0.28$. We integrate position and velocity with respect to time
analytically.
The time step is given by
\begin{eqnarray}
	\Delta t = \frac{\pi}{32\omega}
\end{eqnarray}

In this test, we use first-, third- and fourth-order schemes. The numbers of
particles are $N=33\times33$ and $129\times129$.
We rearrange particles to the grid pattern $n_{\rm rea}$ times within two rotations.
We compare the results with $n_{\rm rea} = 8$, $4$, $2$, $1$ and $0$.
We rearrange at $(16m + 8)$-th, $(32m + 8)$-th, $(64m + 24)$-th and $72$-th steps when the
rotating angle is $\pi/4 + m\pi/2 $ ($m \in \mathbb{N}$) for $n_{\rm rea} = 8$, $4$,
$2$ and $1$.

\begin{figure}[hbtp]
	\centering
	\includegraphics[width = 16cm]{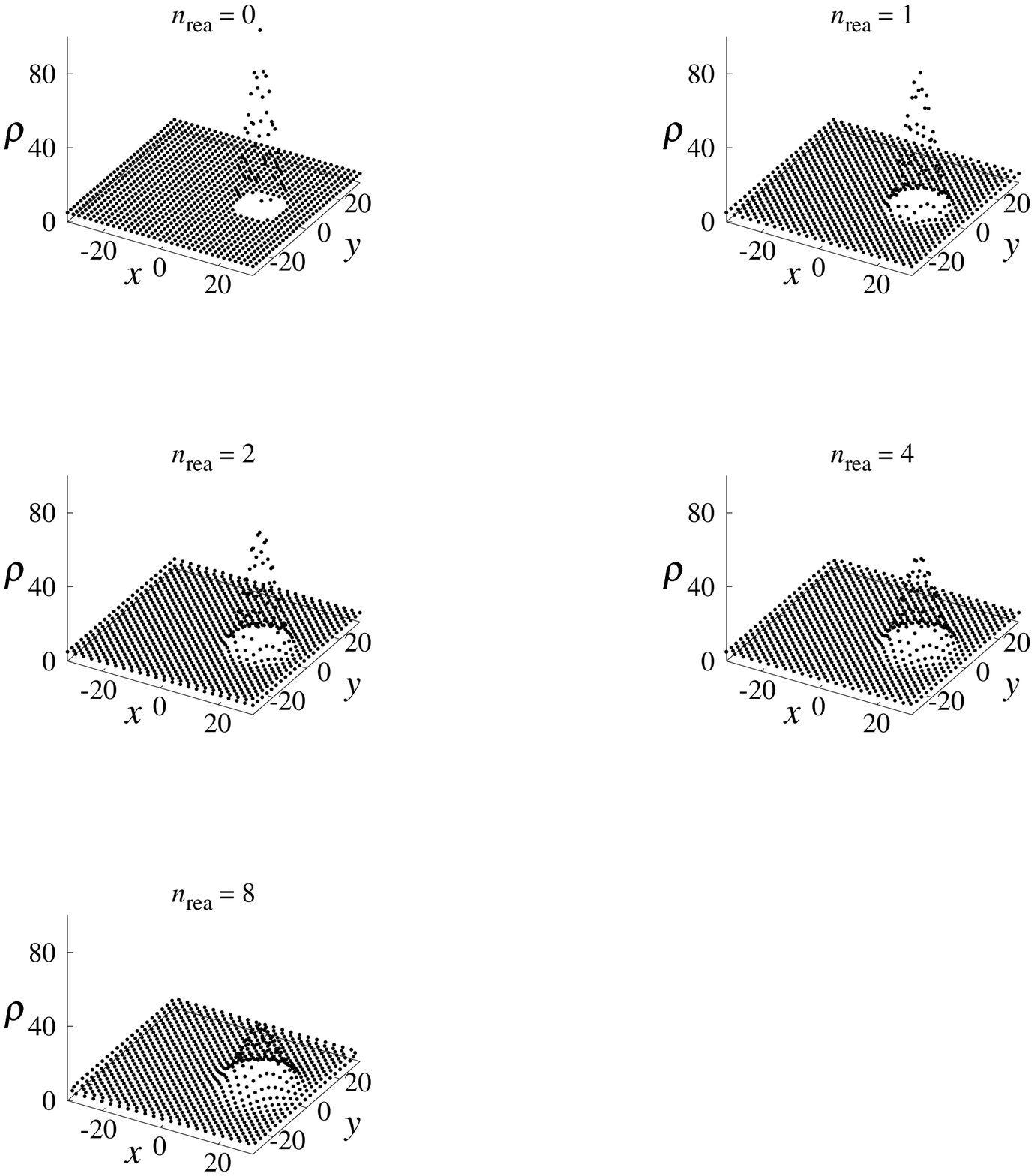}
	\caption{Results of the rotating cone test with $N=33\times33$ and
		the first-order scheme. From left top to right bottom, the
		result without the rearrangement and with $n_{\rm rea} =
		128$, $64$, $32$, $16$ and $8$.}
	\label{fig:rotcone11}
\end{figure}

\begin{figure}[hbtp]
	\centering
	\includegraphics[width = 16cm]{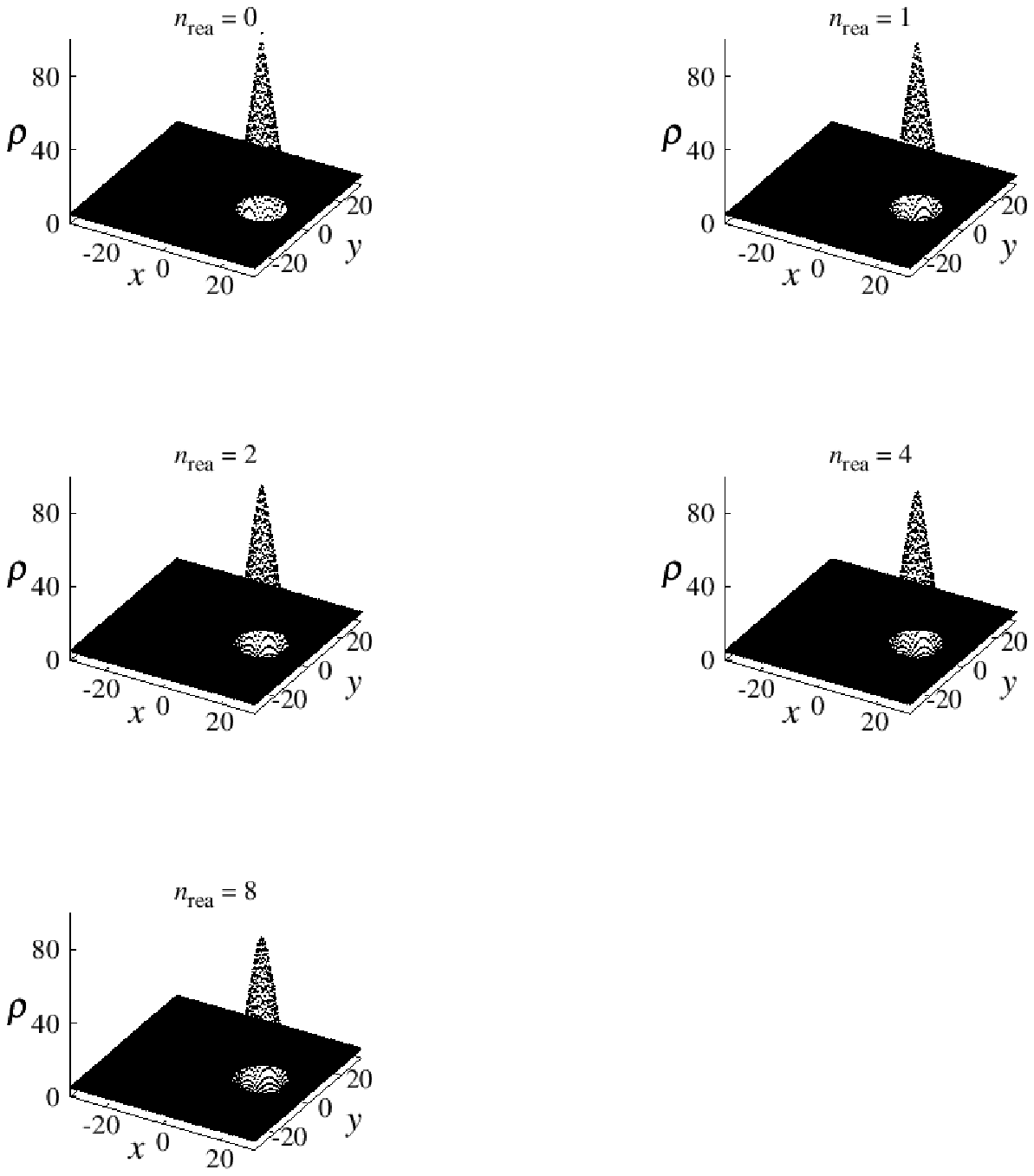}
	\caption{The same as figure \ref{fig:rotcone11}, but for $N = 129\times129$.}
	\label{fig:rotcone14}
\end{figure}

\begin{figure}[hbtp]
	\centering
	\includegraphics[width = 16cm]{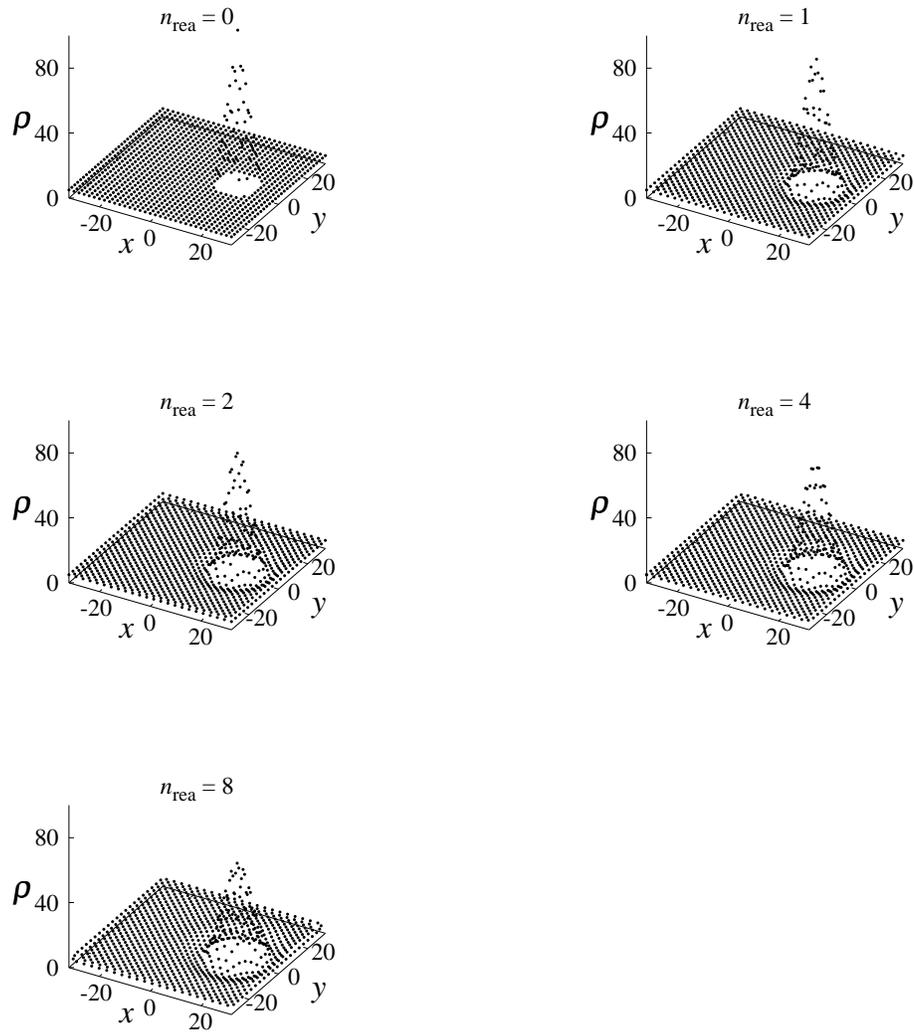}
	\caption{The same as figure \ref{fig:rotcone11}, but for the fourth-order scheme.}
	\label{fig:rotcone41}
\end{figure}

\begin{figure}[hbtp]
	\centering
	\includegraphics[width = 16cm]{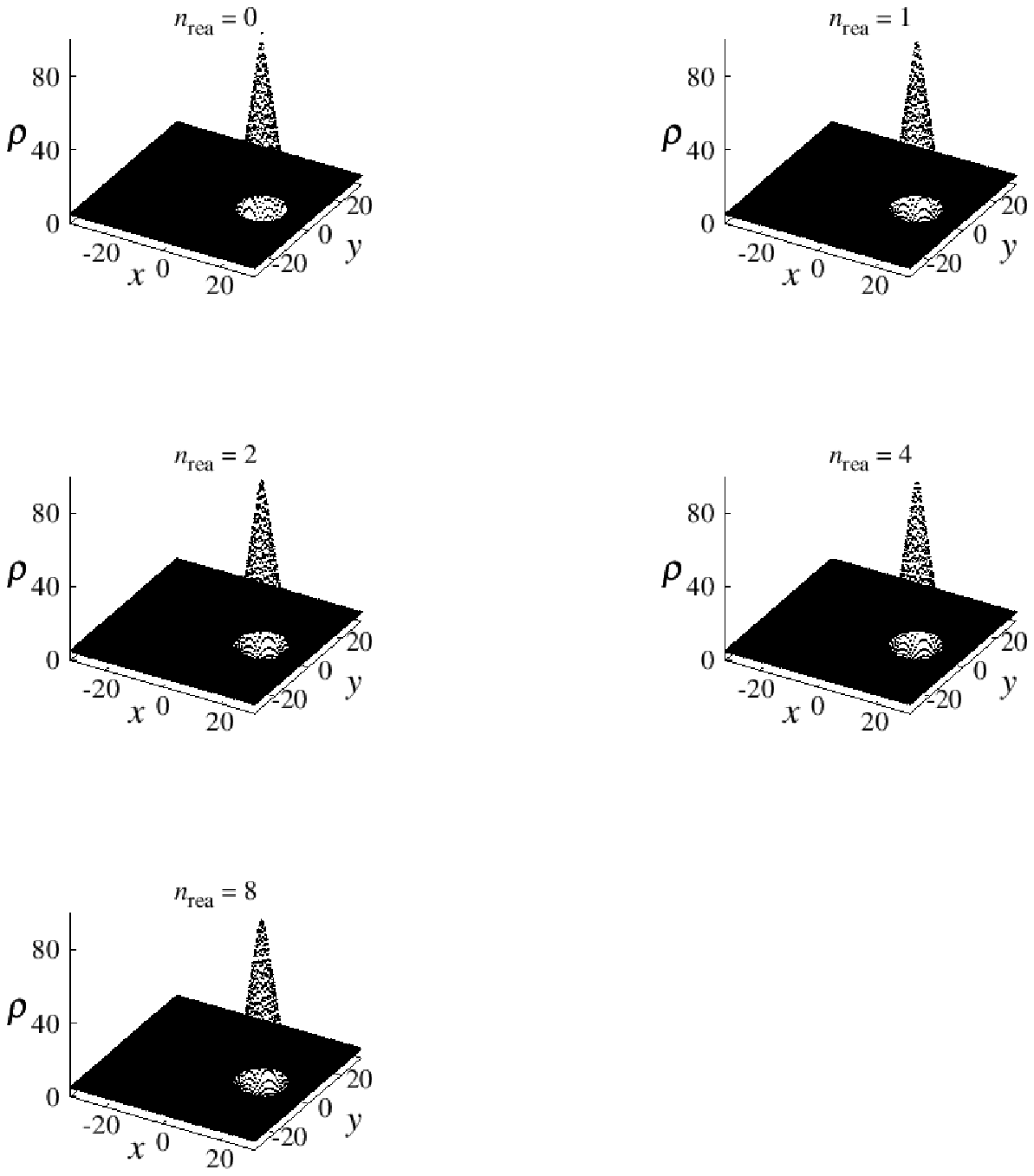}
	\caption{The same as figure \ref{fig:rotcone41}, but for $N = 129\times129$.}
	\label{fig:rotcone44}
\end{figure}

Figures \ref{fig:rotcone11}-\ref{fig:rotcone44} show cones after two rotations
for the number of particles $N=33\times33$ and $N=129\times129$. The orders of
schemes are first and fourth.


Table \ref{tab:rotcone} shows the errors of the height of the cone after
two rotations defined as
\begin{eqnarray}
	\left|1-\frac{\rho_{(x=16, y=0)}}{\rho_{\rm peak}}\right|,
\end{eqnarray}
We derive $\rho_{(x=16, y=0)}$ using equation (\ref{eq:my,sf0}) when $n_{\rm rea}$
is not zero since, due to the rearrangement, there is no particle at $(x,y) = (16,0)$.
We can see that the errors are smaller for larger $n_{\rm rea}$, higher
order schemes, and higher resolution (larger number of particles).
The error for the case of $N = 129 \times 129$ and the fourth-order scheme
are better or similar to those of mesh and mesh-free schemes discussed
in \citet{chock1991comparison} or \citet{vijay1998fluid}, except for those
of extremely high accuracy schemes based on Fourier transform and one of
finite-volume scheme which the interpolynomial is a cubic spline
(\cite{yamartino1993nonnegative}).
In our initial model, we used only one particle to express the peak,
while in the traditional rotating cone test, four grid points are used
to express the peak. Thus, with our initial condition it is much
harder to keep the height of the peak.
Thus, we can conclude that even when we
forced frequent rearrangement of particles, our
scheme can achieve the accuracy comparable to those of high-order
Eulerian schemes. Note that the required frequency of the rearrangement is
generally quite low. For example, the total number of
rearrangement in the Kelvin-Helmholtz instability test with the first-order scheme
discussed in section \ref{sec:test_khi_abel} is 9 in 5291 timesteps.

\begin{table}[htb]
	\tbl{The errors of density peaks.}{
	\begin{tabular}{ccccccc} \hline
			    &     & $n_{\rm rea} = 8$ & $n_{\rm rea} = 4$ & $n_{\rm rea} = 2$ & $n_{\rm rea} = 1$ & $n_{\rm rea} = 0$\\ \hline
		$N = 33\times33$   & 1st & $6.53 \times 10^{-1}$ & $5.19 \times 10^{-1}$ & $4.05 \times 10^{-1}$ &  $3.29 \times 10^{-1}$ & $0.00$ \\
			    & 3rd & $5.32 \times 10^{-1}$ & $4.29 \times 10^{-1}$ & $3.43 \times 10^{-1}$ & $2.86 \times 10^{-1}$ & $0.00$ \\
			    & 4th & $4.07 \times 10^{-1}$ & $3.27 \times 10^{-1}$ & $2.63 \times 10^{-1}$ & $1.89 \times 10^{-1}$ & $0.00$ \\
		$N = 129\times129$ & 1st & $1.79 \times 10^{-1}$ & $1.33 \times 10^{-1}$ & $1.02 \times 10^{-1}$ & $8.26 \times 10^{-2}$ & $0.00$ \\
			    & 3rd & $1.01 \times 10^{-1}$ & $8.49 \times 10^{-2}$ & $7.30 \times 10^{-2}$ & $6.46 \times 10^{-2}$ & $0.00$ \\
			    & 4th & $7.76 \times 10^{-2}$ & $6.85 \times 10^{-2}$ & $6.11 \times 10^{-2}$ & $5.56 \times 10^{-2}$ & $0.00$ \\ \hline
	\end{tabular}}
	\label{tab:rotcone}
\end{table}

\subsection{Kelvin-Helmholtz Instability test}
\label{sec:test_khi}
The Kelvin-Helmholtz instability (hereafter KHI) test has been used to
investigate the ability of numerical schemes to handle large
deformations and hydrodynamical
instabilities (\cite{2003MNRAS.345..429O}, \cite{2007MNRAS.380..963A},
\cite{2008JCoPh.22710040P}, \cite{2012ApJS..201...18M}).
We performed two-dimensional KHI from two different initial conditions.
One is used in \citet{2008JCoPh.22710040P}, and another is in
\citet{2012ApJS..201...18M}.
The difference between two initial conditions is that density and shear
velocity are discontinuous in \citet{2008JCoPh.22710040P}, while smoothed
in \citet{2012ApJS..201...18M}. We present the results of calculations with
the initial condition in \citet{2012ApJS..201...18M} in section
\ref{sec:test_khi_abel} and that in \citet{2008JCoPh.22710040P} in section
\ref{sec:test_khi_price}.

\subsubsection{Kelvin-Helmholtz Instability test with smoothed initial density}
\label{sec:test_khi_abel}
We performed two-dimensional calculations using a computational domain of
$0 \leq x < 1, 0 \leq y < 1$, with a periodic
boundary condition. We made the smoothed
contact discontinuity by setting the initial conditions as
\begin{eqnarray}
	\rho(y) = \left\{
	\begin{array}{ll}
		\rho_l - \rho_m e^{(y-0.25)/\Delta y}, & 0 \leq y < 0.25\\ 
		\rho_h + \rho_m e^{(0.25-y)/\Delta y}, & 0.25 \leq y < 0.5\\ 
		\rho_h + \rho_m e^{(y-0.75)/\Delta y}, & 0.5 \leq y < 0.75\\ 
		\rho_l - \rho_m e^{(0.75-y)/\Delta y}, & 0.75 \leq y < 1
	\end{array}
	\right.
 \end{eqnarray}
where we used $\Delta y = 0.025$, $\rho_h = 2$ and $\rho_l = 1$.
The parameter $\rho_m$ is given by $\rho_m = (\rho_l - \rho_h) /2$.
The smoothed velocity of the $x$-direction (shear velocity) is
 \begin{eqnarray}
	v_x(y) = \left\{
	\begin{array}{ll}
		v_l - v_m e^{(y-0.25)/\Delta y}, & 0 \leq y < 0.25\\ 
		v_h + v_m e^{(0.25-y)/\Delta y}, & 0.25 \leq y < 0.5\\ 
		v_h + v_m e^{(y-0.75)/\Delta y}, & 0.5 \leq y < 0.75\\ 
		v_l - v_m e^{(0.75-y)/\Delta y}, & 0.75 \leq y < 1 
	\end{array}
	\right.
 \end{eqnarray}
where $v_h$ and $v_l$ are the reference values of the $x$-directional
velocity in the high- and low-density regions, respectively. We used
$v_h = -0.5$ and $v_l = 0.5$.
We assume that fluid is an ideal gas with $\gamma = 5/3$ and set
$P = 2.5$ and $v_y = 0$. The number of particles is 256 $\times$ 256.
The velocity perturbation in the $y$-direction is as follows:
 \begin{eqnarray}
	\Delta v_y(x) = A \sin [2\pi x / \lambda],
 \end{eqnarray}
where $\lambda = 0.5$ and $A = 0.01$. The growth timescale of KHI is
 \begin{eqnarray}
	\tau_{\rm KH} = \frac{\lambda(\rho_h + \rho_l)}{\sqrt{\rho_h\rho_l}
	|v_{h} - v_{l}|}.
 \end{eqnarray}
For our setup, $\tau_{\rm KH} = 1.06$.

We rearranged particles with $c_{\rm rea} = 0.55$ for the first-order CPHSF and
$c_{\rm rea} = 0.5$ for the third-order one.

\begin{figure}[ht]
	\begin{center}
	\includegraphics[width=12cm,clip]{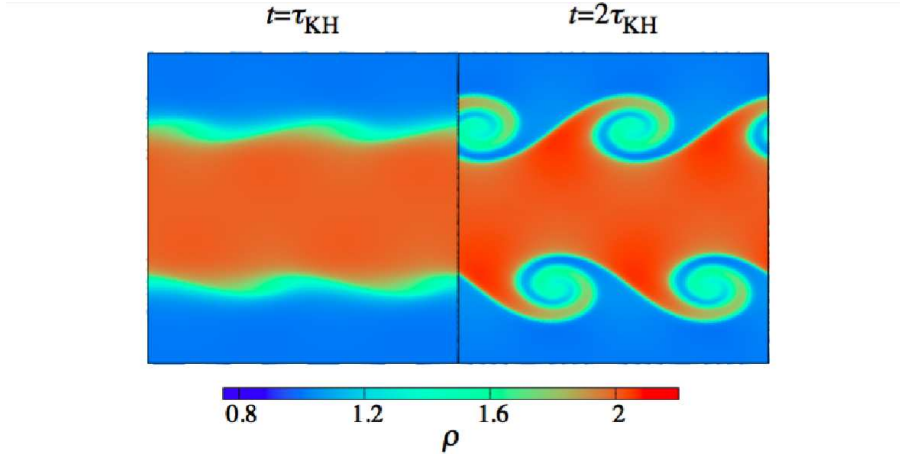} 
	\end{center}
	\caption{Result of the KHI test with smoothed initial density. Density
		distributions at $t= \tau_{\rm kh}$, $2\tau_{\rm kh}$ from
		left to right. The order of the scheme is first.
		}
	\label{fig:khi_mc_1s}
\end{figure}

\begin{figure}[ht]
	\begin{center}
	\includegraphics[width=12cm,clip]{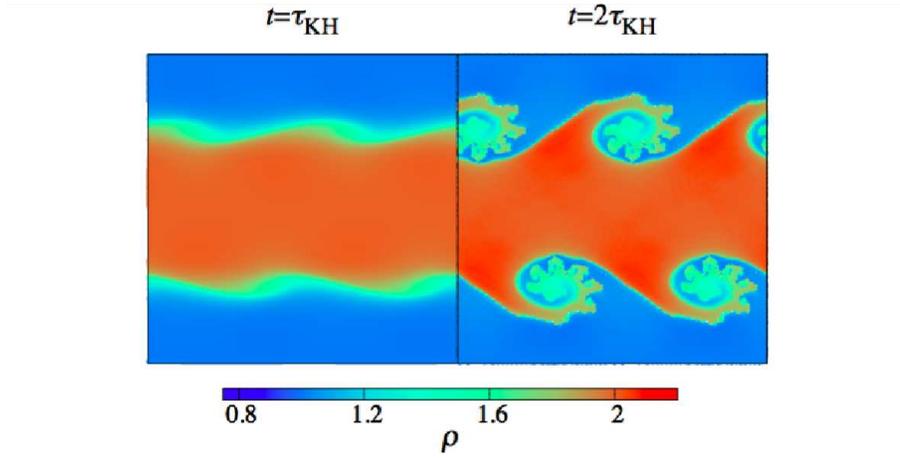}
	\end{center}
	\caption{The same as figure \ref{fig:khi_mc_1s}, but for the
		third-order scheme.}
	\label{fig:khi_mc_3s}
\end{figure}

Figures \ref{fig:khi_mc_1s} and \ref{fig:khi_mc_3s} show the results obtained
using first- and third-order schemes. The first-order result looks similar to
the highest resolution Pencil Code result of \citet{2012ApJS..201...18M}.
In figure \ref{fig:khi_mc_3s},
we can clearly see the development of the second-level KHI, which does not exist in
the first-order result. Thus, we can conclude that the effective resolution of the
third-order scheme is significantly higher than that of the first-order scheme
with the same number of particles.

\subsubsection{Kelvin-Helmholtz Instability test with sharp initial density}
\label{sec:test_khi_price}
The initial condition is same as that in section \ref{sec:test_khi_abel},
except for density, shear velocity and the perturbation. In this simulation,
initial density is given by
\begin{eqnarray}
	\rho(y) = \left\{
	\begin{array}{ll}
		\rho_l, & 0 \leq y < 0.25,\ 0.75 \leq y < 1 \\
		\rho_h, & 0.25 \leq y < 0.75
	\end{array}
	\right.
\end{eqnarray}
where we used $\rho_h = 1$ and $\rho_l= 2$. Velocity of the
$x$-direction is
\begin{eqnarray}
	v_x(y) = \left\{
	\begin{array}{ll}
		v_l, & 0 \leq y < 0.25,\ 0.75 \leq y < 1 \\
		v_h, & 0.25 \leq y < 0.75
	\end{array}
	\right.
\end{eqnarray}
where we used $v_h = -0.5$ and $v_l = 0.5$.
The velocity perturbation in the $y$-direction is
\begin{eqnarray}
	\Delta v_y(x) = \left\{
\begin{array}{ll}
	A \sin [-2\pi (x + 0.5) / \lambda], & 0 \leq |y-0.25| \leq 0.025\\
	A \sin [2\pi (x + 0.5) / \lambda], & 0 \leq |y-0.75| \leq 0.025
\end{array}
\right.
\end{eqnarray}
where $\lambda = 1/6$ and $A = 0.025$. For our setup, $\tau_{\rm KH} = 0.35$.

We rearranged particles with $c_{\rm rea} = 0.55$ for the first-order CPHSF and
$c_{\rm rea} = 0.4$ for the third-order one.

\begin{figure}[ht]
	\begin{center}
	\includegraphics[width=12cm,clip]{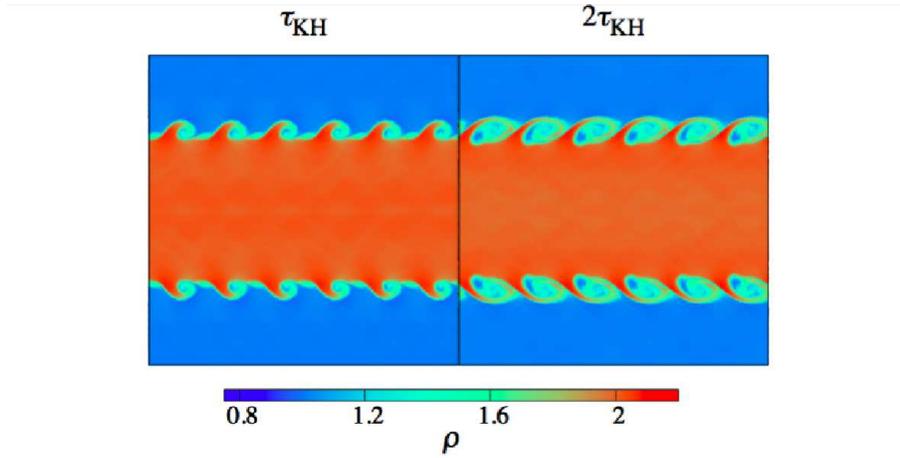}
	\end{center}
	\caption{Result of the KHI test with sharp initial density. Density
		distributions at $t= \tau_{\rm kh}$, $2\tau_{\rm kh}$ from
		left to right. The order of the scheme is first.
		}
	\label{fig:khi_pr_1s}
\end{figure}

\begin{figure}[ht]
 \begin{center}
	\includegraphics[width=12cm,clip]{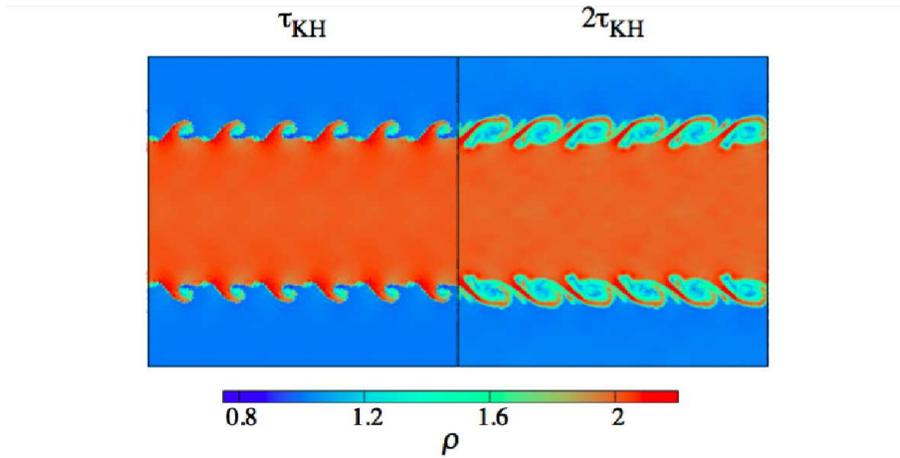}
 \end{center}
	\caption{The same as figure \ref{fig:khi_pr_1s}, but for the
		third-order scheme.}
	\label{fig:khi_pr_3s}
\end{figure}

Figures \ref{fig:khi_pr_1s} and \ref{fig:khi_pr_3s} show the results obtained
using first- and third-order schemes.
The vortexes in our result are clearer than that of \citet{2008JCoPh.22710040P}
in which SPH with the artificial conductivity is used.
We conclude that CPHSF can handle KHI, even if the initial density
and shear velocity at the contact discontinuity are really discontinuous.


\subsection{Rayleigh-Taylor Instability test}
\label{sec:test_rti}
The Rayleigh-Taylor instability (hereafter RTI) test is one of popular tests to
investigate the capability of the scheme to handle fluid instability.
We used the initial condition the same as in \citet{2013ApJ...768...44S}.
The computational domain is $0 \leq x \leq 1$, $0 \leq y \leq 1$, and the
boundary of two fluids is at $y = 0.5$.
We applied a periodic boundary condition at $x = 0$, $1$ and equation
(\ref{eq:boundary,fixedslipping}) at $y = 0$, $1$ as boundary conditions.
The number of particles is $256$
$\times$ $257$. The gravitational acceleration is $-0.5$. Initial velocity is
$v_x = 0$ and $v_y = 0$. Initial density is
\begin{eqnarray}
	\rho(y) = \left\{
\begin{array}{ll}
	\rho_l\left[1+\frac{\gamma-1}{\gamma}\frac{\rho_lg(y-0.5)}{P_0}\right],
	& 0 \leq y \leq 0.5 \\
	\rho_h\left[1+\frac{\gamma-1}{\gamma}\frac{\rho_hg(y-0.5)}{P_0}\right],
	& 0.5 \leq y \leq 1
\end{array}
\right.
\end{eqnarray}
where we used $\rho_h = 2$ and $\rho_l = 1$. Since the fluid is in
equilibrium, initial pressure is given by
 \begin{eqnarray}
	P(y) = \left\{
	\begin{array}{ll}
		P_0 \left(\frac{\rho}{\rho_l}\right)^\gamma,
		& 0 \leq y \leq 0.5 \\
		P_0 \left(\frac{\rho}{\rho_h}\right)^\gamma.
		& 0.5 \leq y \leq 1
	\end{array}
	\right.
 \end{eqnarray}
where $P_0$ is the pressure at the boundary of two fluids, and we used $P_0 =
10/7$ in this test.
The velocity perturbation in the $y$-direction for $0.3 \leq y \leq 0.7$ is
\begin{eqnarray}
	\Delta v_y(x,y) = \delta_{vy}[1+\cos(4\pi x)]
	\{1+\cos[5\pi(y-0.5)]\},
\end{eqnarray}
where we used $\delta_{vy} = 0.025$.

We used equation (\ref{eq:rea,criterion}) with $c_{\rm rea} = 0.55$ for the
rearrangement of particles. Since equation
(\ref{eq:rea,criterion}) cannot be applied to
particles near the surface, we excluded particles with $y < 0.05$ or $y > 0.95$
when evaluating equation (\ref{eq:rea,criterion}).

\begin{figure}[ht]
	\begin{center}
		\includegraphics[width=16cm,clip]{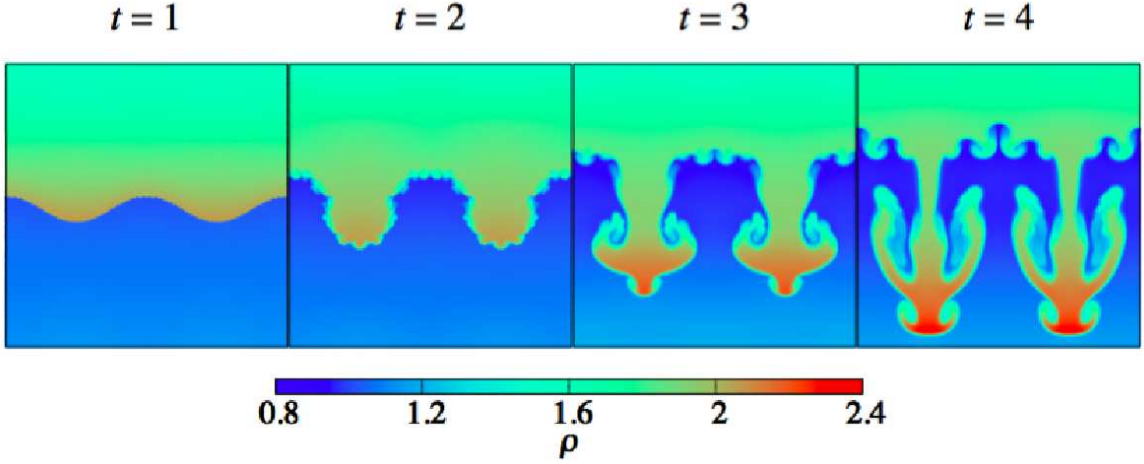} 
	\end{center}
	\caption{Result of the RTI test with sharp initial density.
		Density distributions at $t= 1$, $2$, $3$ and $4$ from
		left to right. The order of the scheme is first.
}
	\label{fig:rti}
\end{figure}

Figure \ref{fig:rti} shows the result. We can see that CPHSF can handle RTI.
We can see small-scale KHI features develop near the bottom of the heavier fluid,
and we can also see small-scale RTI features near the ``root'' of sinking
heavier fluid. These fine features indicate that CPHSF has high resolution and
small dissipation.

\subsection{Gravity wave test}
\label{sec:test_gw}
The gravity wave test is useful to investigate the capability of numerical
schemes to handle the free surface. Note that standard SPH cannot handle
the gravity wave well, because it cannot correctly evaluate the density of
particles near the surface. SPH schemes specially designed to handle free surface
exist (e.g. \cite{1994JCoPh.110..399M}, \cite{2011CoPhC.182..866A}). However,
the most sophisticated schemes require
local high-order diffusion of velocity to stabilize the wave
(e.g. \cite{2011CoPhC.182..866A}).
The computational domain is $0 \leq x < 1$, $0 \leq y < 1$.
We applied a periodic boundary at $x = 0$, equation
(\ref{eq:boundary,freesurface}) at $y = 0$ and equation
(\ref{eq:boundary,fixedslipping}) for particles initially at $y = 1$ as
boundary conditions. The number of particles is $50$ $\times$ $51$.
The equation of state is given by equation (\ref{eq:eos_water})
with $g = -10$, $\rho_{\rm air} = 1000$ and $P_{\rm air} = 10^5$, and initial
density is 
\begin{eqnarray}
	\rho(y) = \rho_{\rm air} e^{g(H-y)/c_{s0}^2}.
\end{eqnarray}
Initial velocity is the same as in \citet{2011CoPhC.182..866A},
\begin{eqnarray}
	v_x
	&=& A\frac{|g|k}{\omega}\frac{\cosh(ky)}{\cosh(kH)}\sin(k x), \\
	v_y
	&=& -A\frac{|g|k}{\omega}\frac{\sinh(ky)}{\cosh(kH)}\cos(k x),
\end{eqnarray}
where $A$, $k$ and $\omega$ are the amplitude, the number of wave and the
frequency. In this test, we set $A = 0.01$, $k = 2\pi / L$ and $\omega =
\sqrt{|g|k\tanh(k H)}$. 
We used $\zeta_{\min} = 0.1$.

\begin{figure}[ht]
	\begin{center}
	\includegraphics[width=12cm,clip]{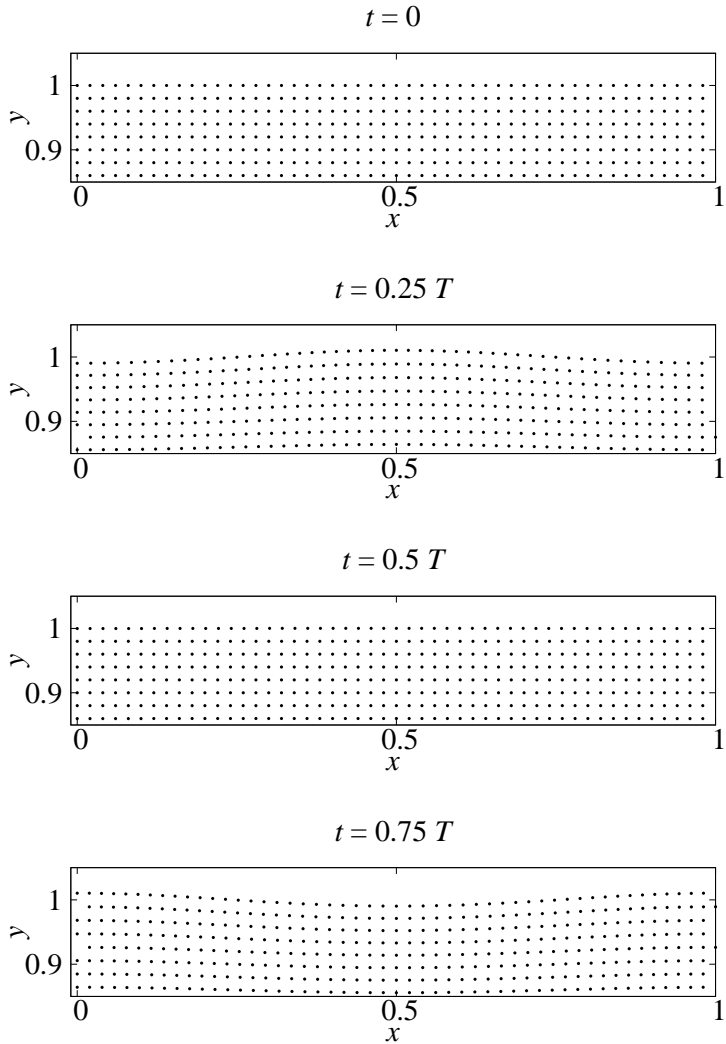}
	\end{center}
	\caption{Results of gravity wave test, from top to bottom, the
		snapshots at $t=0$, $0.25T$, $0.5T$ and $0.75T$ are shown.
		}
	\label{fig:gw}
\end{figure}

\begin{figure}[ht]
	\begin{center}
	\includegraphics[width=12cm,clip]{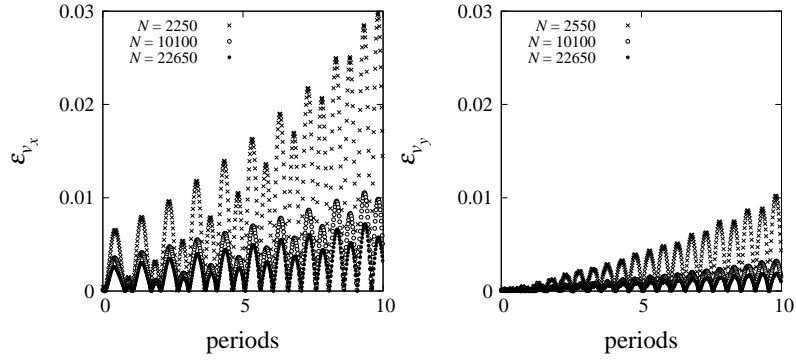} 
	\end{center}
	\caption{The errors of $x$- and $y$-directional
		velocity at $(x, y) = (0.25, 1)$ in the gravity wave test
		with the numbers of particles $N$ $2550$, $10100$,
		$22650$.
		}
	\label{fig:gwe}
\end{figure}

\begin{figure}[ht]
	\begin{center}
	\includegraphics[width=12cm,clip]{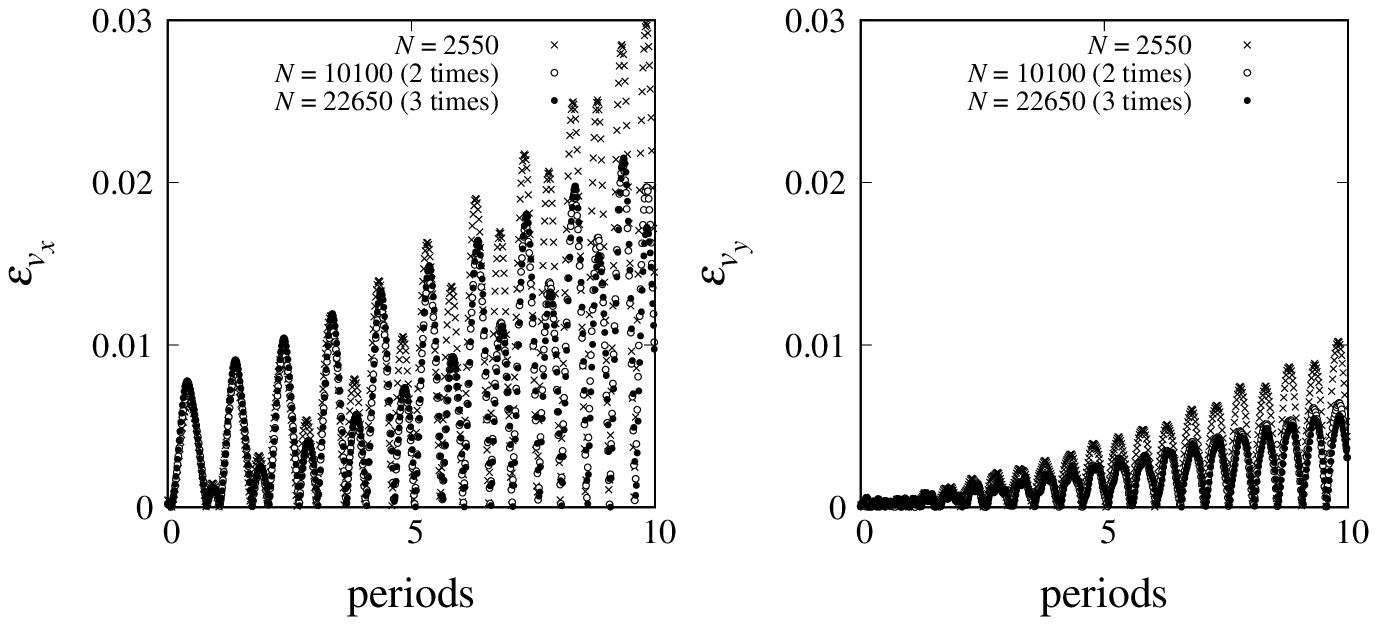}
	\end{center}
	\caption{The same as figure \ref{fig:gwe}, but the error with $N=
		10100$ is multiplied by two, while that with $N=22650$ is
		multiplied by three.}
	\label{fig:gwe2}
\end{figure}

Figure {\ref{fig:gw}} shows the time evolution up to $t = 0.75T$.
Figure {\ref{fig:gwe}} and figure {\ref{fig:gwe2}} show the error of
velocity at $(x, y) = (0.25, 1)$ for runs with the number of particles
$N$, $50 \times 51$,
$100 \times 101$ and $151 \times 150$. The error is given by
\begin{eqnarray}
	\epsilon_{v_x} &=& |v_x - v_{x,0}|,\\
	\epsilon_{v_y} &=& |v_y - v_{y,0}|,
\end{eqnarray}
where $v_{x,0}$ and $v_{y,0}$ are analytical solutions.
We can see that the error becomes smaller as we increase $N$,
showing the first-order convergence, as we used the first-order scheme.

\begin{figure}[ht]
	\begin{center}
	\includegraphics[width=16cm,clip]{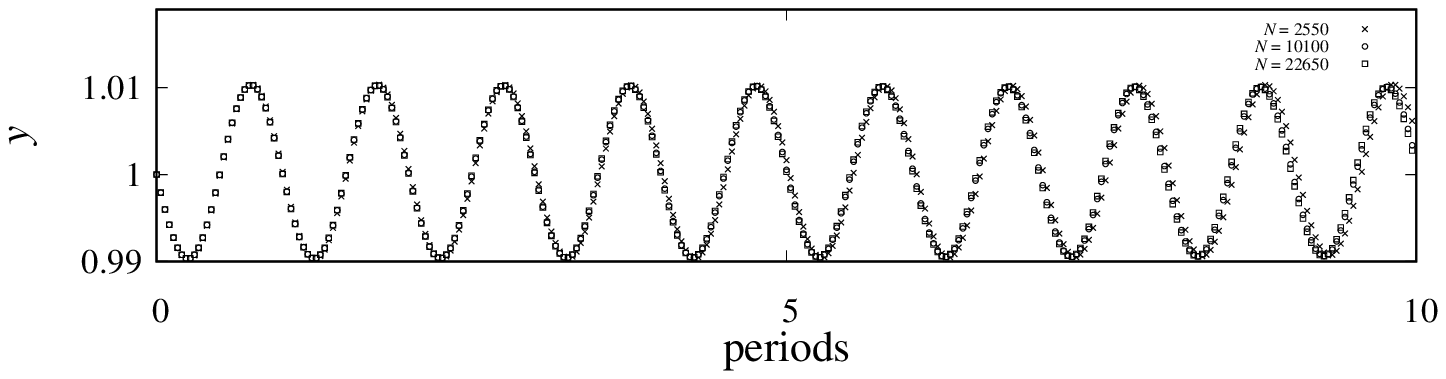}
	\end{center}
	\caption{Time evolution of the $y$-coordinate of the particle
		initially at $(x, y) = (0, 1)$ in the gravity wave test in
		which the numbers of particles $N$ are $2550$, $10100$,
		$22650$.
		}
	\label{fig:gwy}
\end{figure}

Figure {\ref{fig:gwy}} shows $y$ of the particle initially at
$(x, y) = (0, 1)$. In an initial stage, the phases of results for $N=2550$,
$10100$ and $22650$ agree well with each other. However, the wave for
$N = 2550$ has slightly longer period than those of $N=10100$ and $N=22650$, and
the difference of $y$ between the result of $N=2550$ and those of
$N=10100$ and $N=22650$ grows in time. In addition, the phase of
$N=10100$ deviates from that of $N=22650$ slightly at the later stage.

\subsection{Dam break test}
\label{sec:test_dam}
The dam break test is the most widely used test for numerical schemes for the
fluid with the free surface.
The initial condition of the dam break test is the same as that used by
\citet{1994JCoPh.110..399M}. The computational domain is $0 \leq x \leq 0.6$,
$0 \leq y \leq 0.6$. We applied equation
(\ref{eq:boundary,freesurface}) at $x = 0.6$ and $y = 0.6$ and equation
(\ref{eq:boundary,fixedslipping}) at $x = 0$ and $y = 0$ as boundary
conditions. The number of particles is 60 $\times$ 61.
We set that the equation of state is given by equation
(\ref{eq:eos_water}) with
$g = -9.8$, $\rho_{\rm air} = 1000$ and $P_{\rm air} = 10^5$, and initial
density is 
 \begin{eqnarray}
	\rho(y) = \rho_{\rm air} \left[\frac{6|g|}{7C_{B}}\rho_{\rm air}
	(0.6-y) + 1\right]^{1/6}.
 \end{eqnarray}
We set that density and pressure of the particles at
$x = 0.6$ is
\begin{eqnarray}
	\rho &=& \rho_{\rm air},\\
	P &=& P_{\rm air},
\end{eqnarray}
Initial velocity is $v_x = v_y = 0$.

In dam break simulation, the rearrangement of particles is necessary. We
rearrange particles every $0.01$ time unit.
We used $C_{\rm CFL} = 0.05$.

\begin{figure}[ht]
	\begin{center}
	\includegraphics[width=12cm,clip]{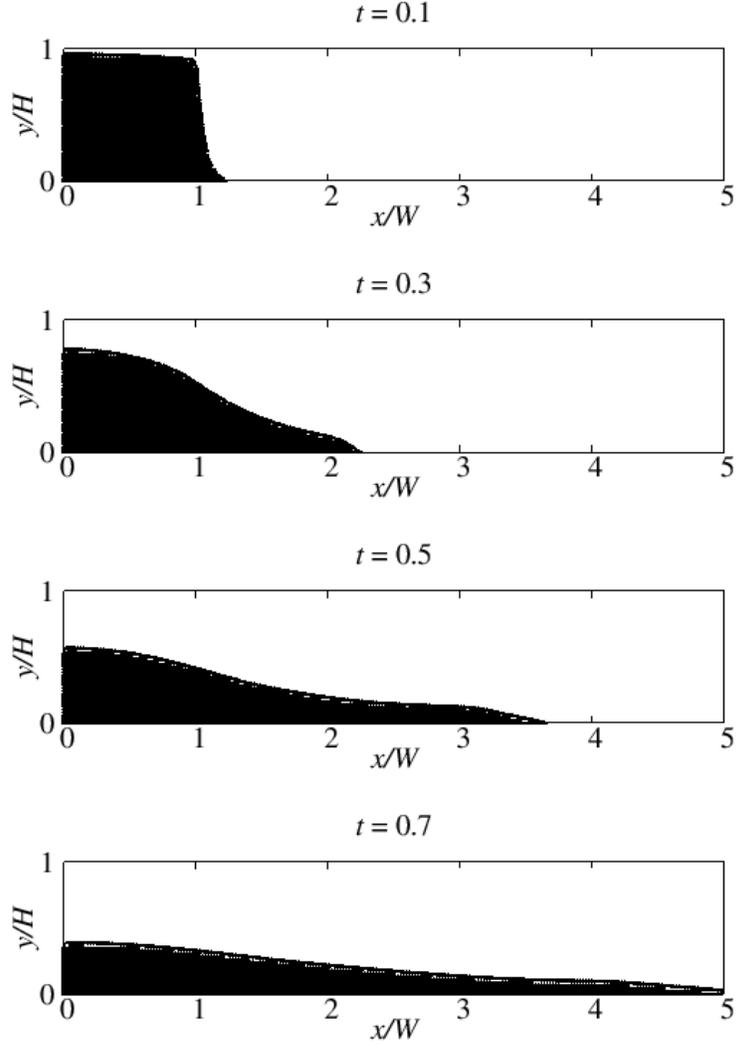}
	\end{center}
	\caption{
		Result of the dam break test.
		The snapshots at $t=0$, $0.25T$, $0.5T$ and $0.75T$ from top to bottom.
	}
	\label{fig:dam}
\end{figure}

\begin{figure}[ht]
	\begin{center}
	\includegraphics[width=6cm,clip]{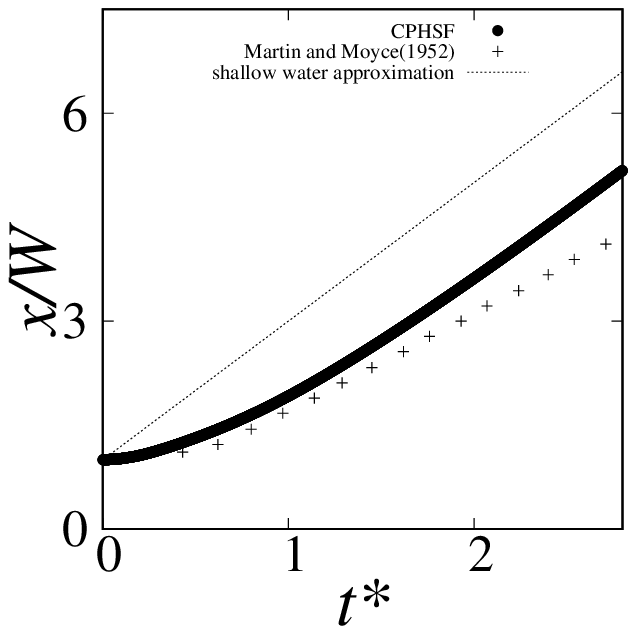} 
	\end{center}
	\caption{The forefront of water in the dam break test plotted
		against $t^*$. Filled circles, crosses show the result of
		first-order CPSHF and experimental result (see text),
		respectively. The solid line indicate the analytic estimate
		based on the shallow water equation.
		}
	\label{fig:damx}
\end{figure}

Figure {\ref{fig:dam}} shows the time evolution up to $t = 0.7$.
The result looks similar to those in the previous dam break test
(\cite{1994JCoPh.110..399M}).
Figure {\ref{fig:damx}} gives the position of the forefront of water
plotted against $t^*$, where $t^*$ is the dimensionless time defined as
$t^* = t\sqrt{|g|/H}$. The results are for first-order CPHSF, an analytical
solution (\cite{whitham1999linear}) and the experimental data of
\citet{1952RSPTA.244..312M}, by \citet{2014JFS....48..407L} distributed
\footnote{http://canal.etsin.upm.es/papers/lobovskyetaljfs2014/}.
It was compared to
numerical solutions in previous studies (e.g. \cite{1994JCoPh.110..399M},
\cite{staroszczyk2010simulation}).
According to \citet{whitham1999linear}, the forefront velocity $v_{\rm dw}$
becomes $v_{\rm dw} = 2\sqrt{|g|H}$ in static state, if we use
on the theory of shallow water waves. Therefore, the position of the
forefront is $2\sqrt{|g|H}t$ analytically.

Note that the analytical solution is not exact. Clearly, in the limit of $t^*
\rightarrow 0$, the solution should be quadratic and not linear in time.
The experimental result is not free from the real viscosity. Therefore,
the discrepancy between our numerical result and experimental result
does not imply the problem in the side of our scheme.
\citet{staroszczyk2010simulation} calculated the dam break using his CSPH and
standard SPH schemes and compared the result with the experimental data.
The agreement between numerical and experimental data was actually pretty good,
better than that in our case. This result probably imply both of experimental result
and \citet{staroszczyk2010simulation}'s result suffer the effect of viscosity
on other dispational effect. Though \citet{staroszczyk2010simulation} did not
explicitly use AV, he applied the re-evaluation of the density, which would cause
significant dissipation. Therefore, we believe our scheme is less dissipative
than his CSPH.

\subsection{Cold Keplerian disk test}
\label{sec:test_disk}
The cold Keplerian disk is important in astronomy.
We prepared a two-dimensional computational domain of $0.5 \leq r < 2$.
We applied equation (\ref{eq:boundary,freesurface}) at the inner edge and
equation (\ref{eq:boundary,fixedslipping}) at the outer edge as boundary
conditions. The number of particles is 46368, and we place the particles
in concentric rings with the same intervals for
$r$ axis and $\theta$ axis. We assumed that the fluid is ideal gas with
$\gamma = 1.4$ and set $\rho = 1$ and $P = 10^{-6}$. Initial velocity is
the pure Keplarian rotation given by
\begin{eqnarray}
	v_x &=& -\sqrt{\frac{GM_{\ast}}{r}}\frac{y}{r}, \\
	v_y &=& \sqrt{\frac{GM_{\ast}}{r}}\frac{x}{r},
\end{eqnarray}
where $G$ is the gravitational constant and $M_{\ast}$ is the mass of the
central star. We set $GM_{\ast} = 1$.
The time step is smaller one between the time step given by the Courant
conditions for the pressure gradient term and the central force term.
Therefore, we compare equation (\ref{eq:dt_cfl}) with
\begin{eqnarray}
	\Delta t = \min_i c_{\rm disk}\sqrt{\frac{r_i}{\|\boldsymbol{a}_i\|}},
\end{eqnarray}
and then smaller one is taken. The parameter $c_{\rm disk}$ is a constant
coefficient, and the value of $2\pi/c_{\rm disk}$ denotes how many steps is
calculated in a orbit. We set $2\pi/c_{\rm disk} = 50$.

In this test, we used fourth-order CPHSF, because the accuracy of the
approximation of the artificial viscosity is important to prevent unphysical
angular momentum transfer (e.g. \cite{2010MNRAS.408..669C}).
In addition, we use fifth-order Radau method for the time integration.

\begin{figure}[ht]
	\begin{center}
		\includegraphics[width=16cm,clip]{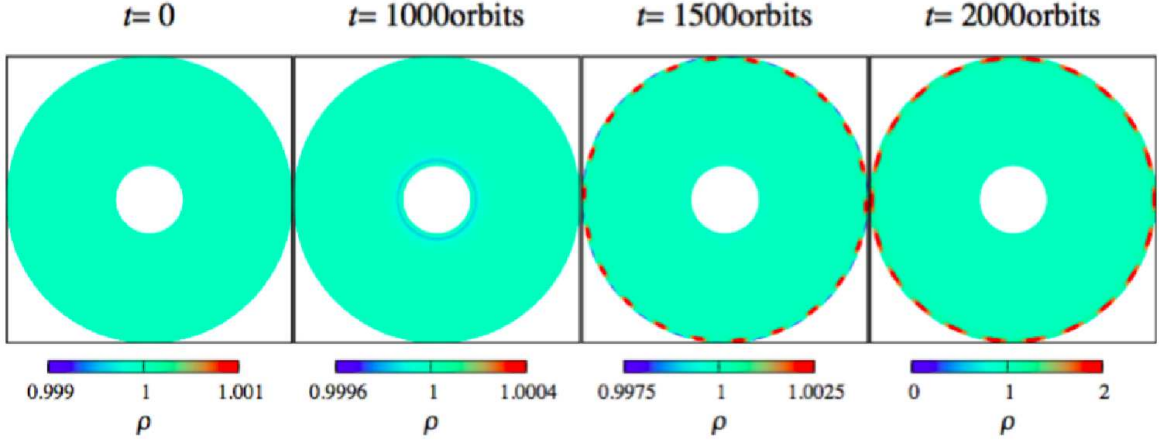} 
	\end{center}
	\caption{The result of the cold Keplerian disk test. 
		Density distributions at $t\simeq$ $0$, $1000$,
		$1500$ and $2000$ orbits from left to right.
		}
	\label{fig:disk}
\end{figure}

Figure {\ref{fig:disk}} shows the results. We can see that the disk does
not break till $t\simeq 2000$ orbits. Note that this lifetime is longer
than the lifetime of any Lagrangian scheme, reported
in the literature (\cite{2015MNRAS.450...53H}, \cite{2016ApJS..224...32H})
by more than a factor three.
	
\begin{figure}[ht]
	\begin{center}
		\includegraphics[width=6cm,clip]{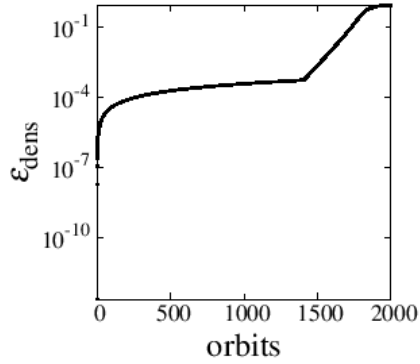} 
	\end{center}
	\caption{The maximum error of density in particles plotted against
		the number of orbits in the cold Keplerian disk test.}
	\label{fig:diske}
\end{figure}

Figure \ref{fig:diske} shows the error of the density plotted against time.
The error is given by
 \begin{eqnarray}
	\epsilon_{\rm dens} = \max_i \left| \rho_i - 1 \right|.
 \end{eqnarray}
It is clear that the error grows rapidly after $t=1400$ orbits. From figure
{\ref{fig:disk}}, we can see that the error becomes large at the
outer edge after $t = 1500$ orbits. Therefore, if we can improve the
treatment of the outer edge, we may be able to reduce the error. We conclude
that the cold Keplerian disk integrated using CPHSF can survive for $> 1000$
orbits.
CPHSF can handle rotating disks better than the Lagrangian scheme variants
proposed in previous studies.

\section{Discussion and summary}
\label{sec:sumanddis}
\subsection{Discussion}
One limitation of our current CPHSF scheme is that there is no easy way to
handle topological changes of fluid, for example the collision of two droplets.
In principle, we can handle collisions by detecting the moment of the collision
and change the surface particles (usually with special boundary conditions)
to bulk particles. However, finding the exact moment of collision for each of
surface particles involved can be very expensive. Therefore approximate
treatment is practically necessary.
Another problem is that we currently need implicit time integration in order to
apply boundary conditions. An explicit or semi-implicit scheme would be
better, and should be applicable at least to the bulk of the fluid.

\subsection{Summary}
Standard SPH scheme and its variations are not high order and cannot handle
free surfaces. We formulate CPHS, a high-order mesh-free method.
CPHSF can handle Kelvin-Helmholtz instability and Rayleigh-Taylor
instability better than previous SPH methods do. We also simulate the gravity
wave and dam break. The results are excellent. Additionally, the cold
Keplerian disk, which is very important in astrophysics,
can be calculated for much larger time than possible with previous mesh-free methods.

CPHSF does not exactly satisfy the conservation laws. However we showed that
the violation of the conservation laws is acceptable, because of the Lagrangian
nature and high-order accuracy of CPHSF.
Therefore we conclude that CPHSF is more useful than previous mesh-free method.


\bigskip

\begin{ack}
We thanks Takayuki Saitoh for discussion about mesh-free method and others,
Natsuki Hosono, Masaki Iwasawa and many others for providing useful comments
about numerical simulations, Keigo Nitadori for discussion about formulation.
We thank the referee for insightful and helpful comments.
This work was supported by RIKEN Junior Research Associate Program and MEXT
SPIRE and JICFuS. Part of the research covered in this paper was funded by
MEXT's program for the Development and Improvement for the Next Generation
Ultra High-Speed Computer System, under its Subsidies for Operating the
Specific Advanced Large Research Facilities.
\end{ack}




\end{document}